\documentclass{jpp}
\usepackage[T1]{fontenc}
\usepackage[utf8]{inputenc}
\usepackage{xcolor}
\usepackage{units}
\usepackage{mathrsfs}
\usepackage{mathtools} 	
\usepackage{url}
\usepackage{amsmath}
\usepackage{amssymb}
\usepackage[pdfusetitle,
 bookmarks=true,bookmarksnumbered=false,bookmarksopen=false,
 breaklinks=true,pdfborder={0 0 0},pdfborderstyle={},backref=false,colorlinks=true]
 {hyperref}
\hypersetup{
 citecolor=blue, linkcolor=blue}

\allowdisplaybreaks

\global\long\def\pFq#1#2{\prescript{\vphantom{#1}}{#1}{F}^{\vphantom{#2}}_{#2}}%

\title{The dielectric tensor for a fully anisotropic regularized Kappa plasma}
\date{\date{}}
\author{Rudi Gaelzer\aff{1,2}
			\corresp{\email{rudi.gaelzer@ufrgs.br}},
Horst Fichtner\aff{2}, Dustin Lee Schröder\aff{2}, Marian Lazar\aff{3},
\and Klaus Scherer\aff{2}}

\affiliation{\aff{1}Instituto de Física, Universidade Federal do Rio Grande do Sul, Brazil
\aff{2}Institut für Theoretische Physik IV, Ruhr-Universität Bochum, Germany
\aff{3}Centre for Mathematical Plasma Astrophysics, KU Leuven, Belgium}

\begin{document}

\maketitle

\begin{abstract}
It has been known for some decades that the plasma in the Solar System is
invariably in a state away from thermal equilibrium. In several plasma
environments, be it in the Solar Wind or within planetary magnetospheres, the
plasma has been found in a quasi-stationary nonthermal state, with a substantial
fraction of its high-energy particles fitting along power-law distributions,
instead of the exponential decay of thermal systems.
Translating the observations in energy to velocity distribution
functions (VDFs), the Kappa plasma model has been employed by the space plasma
community, with the (standard) Kappa VDF (SKD) being the preferred choice.
However, recent works have called attention to the limitations of the SKD model,
proposing as a viable alternative the regularized Kappa VDF (RKD) model, which
corrects the deficiencies of the SKD\@. The RKD model introduces a Gaussian-like
dependency on the very high-energy part of the VDF that effectively cut-offs,
via a regularization parameter, the Lorentzian profile of the
intermediate-energy range. In this work, we employ a generalized RKD that
supports anisotropies both in the thermal speed parameters and in the
regularization parameters, including as well the possibility of drifts parallel
to the magnetic field. This generalized RKD (called the bi-regularized Kappa
with drift distribution, bRKd), is employed to derive for the first time the
components of the susceptibility tensor for a suprathermal plasma
species/population modelled by the bRKd. The tensor components are valid for
waves with any propagation angle, frequency range and polarization. During the
derivation, new special functions are defined and analyzed, which generalize
previous expressions restricted to the parallel propagation limit.
\end{abstract}

\section{Introduction}
The heliosphere, in general, and the solar wind, in particular, remain important natural laboratories for the in-situ observation of plasmas out of thermal equilibrium, see, e.g., the recent reviews by \citet{Fraternale-etal-2022}, \citet{Verscharen-etal-2019, Verscharen-etal-2022}, \citet{Cranmer-etal-2023}, \citet{Zhao-etal-2025}. Within such, mostly collision-poor, plasmas waves play a fundamental role: they determine the level of turbulence, a variety of transport process, reaching from the momentum and energy exchange between (supra)thermal plasma populations, the heat conduction within those, and their acceleration and heating, via the advective and diffusive transport of energetic particles, to the coupling of different scales as well as to the relaxation of phase space distributions. 

While the evolution of the latter can be treated self-consistently with nonlinear modeling \citep[e.g.,][]{Yogesh-etal-2025, Lopez-etal-2026} or, often to a good approximation, within quasi-linear theory \citep[e.g.][]{Walters-etal-2023, Moya-etal-2026}, linear kinetic theory remains a highly useful and frequently employed approach to study the growth, propagation as well as damping of plasma waves in thermal and suprathermal plasmas \citep[e.g.][and references therein]{Yoon-2017, Schroeder+26_JPP}. A central element of the linear theory of plasma waves is the dielectric tensor, in which the plasma properties are encoded via the velocity distribution functions (VDFs) of its populations and the large-scale magnetic field. It defines the structure of the corresponding dispersion relation, whose solutions are all possible plasma modes and their characteristics. The general form of the dielectric tensor for unspecified distribution functions has long been worked out and can be found in various monographs \citep[e.g.,][]{Klimontovich-1983, Stix-1992, Brambilla98, Swanson-2003}. 

Over the years, numerical codes have been developed that allow to solve the general dispersion relation for (gyrotropic but otherwise) arbitrary VDFs and waves with any propagation angle, frequency range as well as polarization \citep[][]{Matsuda-Smith-1992, Astfalk-Jenko-2017, Verscharen-etal-2018}. While these codes have been, and continue to be, essential for numerous studies of waves in non-Maxwellian plasmas, they have principal limitations. First, in general, obtaining specific solutions is numerically costly and, consequently, time-consuming. Second, obviously they do not allow to systematically derive (approximate) analytical expressions for the growth and damping rates as well as the corresponding dispersion curves. The first limitation has been addressed by developing numerical codes for specified VDFs, like WHAMP \citep{Roennmark-1982}, NHDS \citep{Verscharen-Chandran-2018}, or PLUME \citep{Klein-etal-2025} for (bi)Maxwellian distributions and DSHARK \citep{Astfalk-etal-2015} or DIS-K \citep{Lopez-etal-2021} for standard (bi)kappa distributions (SKDs). Given that this strategy is based on an analytical exploitation of a prescribed VDF to formulate the dielectric tensor and, thus, the dispersion relation more explicitly before its numerical solution, it is considered as a semi-analytical approach \citep[see, e.g.,][]{Gaelzer+26/06-arxiv, Tischmann-etal-2026}. To overcome the second limitation, often the dispersion relation is analytically formulated in terms of the corresponding plasma dispersion function, for which asymptotic expansions can be derived, see the overview given in \citet{Gaelzer+24/07}. 

The outlined programme, i.e.\ starting with a fully-numerical approach \citep{Schroeder+25/03, Schroeder+25/07, Schroeder+26_JPP, Schroeder+26/04} followed by a semi-analytical one \citep{Gaelzer+24/07, Gaelzer+26/06-arxiv} to eventually derive analytical approximations to dispersion curves and growth and damping rates, has been performed to some extent also for the so-called regularized (bi-)kappa distribution (RKD) that gained some popularity in recent years, because it improves the SKD approach by removing a number of physical deficiencies \citep[see, e.g., the discussion in][]{LazarFichtner21}. The mentioned semi-analytical approaches to waves in RKD plasmas were, so far, limited to parallel propagation. With the present paper we remove this limitation by deriving the dielectric tensor for arbitrary wave propagation in a fully anisotropic RKD plasma. This way we supplement previous explicit formulations of the dielectric tensor provided for (bi-)Maxwellian plasmas \citep{Gary-1993}, cold relativistic plasmas \citep{Bret-2009}, and (bi-)SKD plasmas \citep{Gaelzer+16/06} to lay the foundation for future improved (semi-)analytical studies of anisotropic RKD plasmas.

The paper is organized as follows. In section~2 the (generalized) fully anisotropic RKD is introduced and the corresponding dielectric and susceptibility tensors are derived. Section~3 contains an analysis of the special functions naturally appearing in the formulation of these tensors. This particularly comprises new anisotropic RKD plasma dispersion functions. All results are summarized in secion~4.

\section{The distribution function and the dielectric tensor}

In this section, the formal expression for the susceptibility tensor of a regularized kappa distribution with drift will be derived.  We will start by presenting the velocity distribution and then, from the general expression of the dielectric tensor of a magnetized plasma, the susceptibility tensor and some of its important limiting cases will be derived.

\subsection{The fully anisotropic regularized distribution function with drift}

The model VDF employed in this work is the same adopted by \citet{Gaelzer+26/06-arxiv},
which extends both the generalized form introduced by \citet{Scherer+20/07,Gaelzer+24/07},
and the drift-anisotropic distribution employed by \citet{Schroeder+25/07,Schroeder+26/04}\@.
Namely,
\begin{equation}
f^{(\eta,\zeta,\boldsymbol{\mu})}_{\kappa,\boldsymbol{\theta}}\left(v_{\parallel},v_{\perp}\right)=N^{(\eta,\zeta,\boldsymbol{\mu})}_{\kappa,\boldsymbol{\theta}}\left[1+\frac{\left(v_{\parallel}-U_{\parallel}\right)^{2}}{\eta\theta^{2}_{\parallel}}+\frac{v^{2}_{\perp}}{\eta\theta^{2}_{\perp}}\right]^{-\zeta}\exp\left[-\frac{\mu_{\parallel}}{\theta^{2}_{\parallel}}\left(v_{\parallel}-U_{\parallel}\right)^{2}-\frac{\mu_{\perp}}{\theta^{2}_{\perp}}v^{2}_{\perp}\right],\label{eq:RKAP3:RKD-anisotropic-beam-generalized}
\end{equation}
where $\boldsymbol{\theta}=\left(\theta_{\parallel},\theta_{\perp}\right)$,
$\boldsymbol{\mu}=\left(\mu_{\parallel},\mu_{\perp}\right)$ and $N^{(\eta,\zeta,\boldsymbol{\mu})}_{\kappa,\boldsymbol{\theta}}$
is the normalization constant, which is evaluated from the usual condition
$\int d^{3}v\,f\left(v_{\parallel},v_{\perp}\right)=1$ and is given
by Eq. (\ref{eq:Norm_const})\@. In (\ref{eq:RKAP3:RKD-anisotropic-beam-generalized}),
parameters $\eta$ and $\zeta$ are coupled via the $\kappa$ parameter,
and $\mu_{\parallel(\perp)}$ are nonnegative dimensionless constants,
called regularization parameters, which do not necessarily depend
on $\kappa$ \citep{Scherer+17/12,Husidic+20/04,Scherer+22/07}\@.
Their values balance the reduction of unphysical effects due to superluminal
particles with the need to retain suprathermal populations in the
high-energy tails \citep{Scherer+19/08} and at the same time enable
the evaluation of all moments of the distribution \citep{Scherer+17/12,Scherer+20/07,Husidic+22/03}\@.
The anisotropic parameters $\theta_{\parallel(\perp)}$ have the dimension
of speed and measure the statistical spread of velocities of the particles
in the parallel (perpendicular) direction relative to the ambient
magnetic field. For this reason, they determine the kinetic measures
of temperature in both directions. 

The VDF $f^{(\eta,\zeta,\boldsymbol{\mu})}_{\kappa,\boldsymbol{\theta}}\left(v_{\parallel},v_{\perp}\right)$
can reproduce conditions that give rise to several different kinds
of instabilities in a plasma that contains suprathermal particles,
such as temperature-anisotropy instabilities (due to $\theta_{\parallel}\neq\theta_{\perp}$)
and beaming, heat-flux instabilities (due to $U_{\parallel}\neq0$)\@.
The support of the additional freedom of choosing $\mu_{\parallel}\neq\mu_{\parallel}$
allows the description of a wider range of particle anisotropies,
while at the same time keeping in effect the important task performed
by the regularization parameter in a suprathermal plasma. This additional
freedom will consequently alter the characteristics of wave propagation
and wave-particle interaction in such a plasma. Initial analyses of
these effects in parallel-propagating waves in typical solar wind
plasmas were conducted by \citet{Gaelzer+26/06-arxiv}·

The relevance of the choice of an anisotropic cut-off is not so much practical but rather closer to reality.
The existence of such cut-offs can be justified by the fact that,
due to the finite time and spatial scales of a given physical system,
any heating or acceleration process has limited effects, thereby constraining the energy ranges or levels of the plasma populations in question. 
Since these processes 
occur in different scales, they
are in general different along and perpendicular to the local magnetic background field, 
and consequently
the corresponding energy limits determining the cutoffs of the VDFs are also expected to be unequal. 
While not modelled in the present study, even the form of the cut-off, here chosen to be exponential, may be different, like, e.g., in the case of a truncation of a distribution to particle escape. 
Therefore, the assumption of an isotropic cutoff appears to be a rather special choice.%



The distribution $f^{(\eta,\zeta,\boldsymbol{\mu})}_{\kappa,\boldsymbol{\theta}}\left(v_{\parallel},v_{\perp}\right)$
will be referred here as the bi-regularized Kappa distribution with
drift (bRKd) and it generalizes several models of anisotropic suprathermal
VDFs employed in the literature. When $\mu_{\perp}=\mu_{\parallel}=\mu$,
it reduces to the drifting anisotropic RKDs employed by \citet{Schroeder+25/07,Schroeder+26/04}
(with $\mu=\alpha^{2}$), where comprehensive analyses of heat-flux
instabilities are performed. This is the case we will refer here as
the $\mu$-isotropic limit. Additional analyses of wave instabilities
arising solely from temperature anisotropies (with $U_{\parallel}=0$)
were performed by \citet{Schroeder+25/03,Schroeder+26_JPP}. In the
fully isotropic case, with $\theta_{\parallel}=\theta_{\perp}=\theta$,
distribution (\ref{eq:RKAP3:RKD-anisotropic-beam-generalized}) reduces
to the model employed by \citet{Gaelzer+24/07} in the analysis the
propagation and absorption of Langmuir waves.

Further particular cases of $f^{(\eta,\zeta,\boldsymbol{\mu})}_{\kappa,\boldsymbol{\theta}}\left(v_{\parallel},v_{\perp}\right)$
include the limit $\mu_{\perp}=\mu_{\parallel}=0$, in which case
it reduces to the standard Kappa VDF given by 
\[
f^{(\eta,\zeta,0)}_{\kappa,\boldsymbol{\theta}}\left(v_{\parallel},v_{\perp}\right)=N^{(\eta,\zeta,0)}_{\kappa,\boldsymbol{\theta}}\left[1+\frac{\left(v_{\parallel}-U_{\parallel}\right)^{2}}{\eta\theta^{2}_{\parallel}}+\frac{v^{2}_{\perp}}{\eta\theta^{2}_{\perp}}\right]^{-\zeta},
\]
where $N^{(\eta,\zeta,0)}_{\kappa,\boldsymbol{\theta}}$ is now given
by (\ref{eq:Norm_const-SKD})\@. This distribution will be referred
to here as the bi-(standard) Kappa distribution with drift (bSKd)
and it contains several models employed in the literature \citep{Livadiotis17,LazarFichtner21}
that can be chosen with appropriate values of the parameters $\left\{ \eta,\zeta,\theta_{\parallel},\theta_{\perp}\right\} $\@.
Here, whenever the limit $\mu=0$ is considered, we will refer to
the generalized bSKd model introduced by \citet{GaelzerZiebell16/02,Gaelzer+16/06},
\begin{equation}
f^{\left(\alpha\right)}_{\kappa,\boldsymbol{w}}\left(v_{\parallel},v_{\perp}\right)=\frac{\left(\pi\kappa\right)^{-3/2}\Gamma\left(\sigma\right)}{w_{\parallel}w^{2}_{\perp}\Gamma\left(\sigma-\nicefrac{3}{2}\right)}\left[1+\frac{\left(v_{\parallel}-U_{\parallel}\right)^{2}}{\kappa w^{2}_{\parallel}}+\frac{v^{2}_{\perp}}{\kappa w^{2}_{\perp}}\right]^{-\sigma},\qquad\left(\sigma>\frac{3}{2}\right),\label{eq:KAP7:f_s}
\end{equation}
where $\Gamma\left(z\right)$ is the Gamma function \citep{AskeyRoy-Full-NIST10}
and $\sigma=\kappa+\alpha$\@. The transition $f^{(\eta,\zeta,0)}_{\kappa,\boldsymbol{\theta}}\longrightarrow f^{\left(\alpha\right)}_{\kappa,\boldsymbol{w}}$
is affected by the recipe $\eta\to\kappa$, $\zeta\to\sigma$ and
$\boldsymbol{\theta}\to\boldsymbol{w}$\@. The final choice $\alpha=1$
reproduces the bSKd model employed by the DIS-K dispersion solver
\citep{Lopez+21/06}.

Finally, taking the limit $\kappa\to\infty$ (with $w_{\parallel}\to v_{T\parallel}$
and $w_{\perp}\to v_{T\perp}$), we obtain the bi-Maxwellian with
drift (bMd) distribution,
\[
f_{\mathrm{bMd}}\left(v_{\parallel},v_{\perp}\right)=\frac{1}{\pi^{3/2}v_{T\parallel}v^{2}_{T\perp}}\exp\left(-\frac{\left(v_{\parallel}-U_{\parallel}\right)^{2}}{v^{2}_{T\parallel}}-\frac{v^{2}_{\perp}}{v^{2}_{T\perp}}\right),
\]
where now $v_{T\parallel(\perp)}=\sqrt{2T_{\parallel(\perp)}/m}$
is the parallel (perpendicular) thermal speed.

\subsection{Evaluation of the dielectric tensor}

\begin{subequations}
\label{eq:General_dielectric_tensor}

Closed-form expressions for the components of the full dielectric
tensor of (standard) suprathermal plasmas were derived by \citet{GaelzerZiebell16/02,Gaelzer+16/06,Lopez+21/06}\@.
Here, we will derive for the first time the full tensor for a bRKd
plasma. The general form of the Cartesian components of the dielectric
tensor for a gyrotropic, homogenous magnetized plasma can be written
as \citep{Brambilla98}
\begin{equation}
\varepsilon_{ij}\left(\boldsymbol{k},\omega\right)=\delta_{ij}+\sum_{s}\chi^{(s)}_{ij}\left(\boldsymbol{k},\omega\right),\label{eq:GDT-Dielectric_Tensor}
\end{equation}
where $i,j=1,2,3$ (or $x,y,z$), $\boldsymbol{k}$ is the wave vector
and $\omega$ is the (angular) frequency of the wave. Here, $\chi^{(s)}_{ij}\left(\boldsymbol{k},\omega\right)$
is the $ij$ component of the partial susceptibility tensor that contains
the contributions to $\varepsilon_{ij}$ from particles of species/population
$s$ $\left(s=e,p,\dots\right)$\@.

The partial susceptibility of the $s$-population is then given by
\begin{equation}
\chi^{(s)}_{ij}\left(\boldsymbol{k},\omega\right)=\sum^{\infty}_{n\to-\infty}\chi^{(s)}_{n,ij}\left(\boldsymbol{k},\omega\right)+\delta_{iz}\delta_{jz}\tilde{\chi}^{(s)}_{zz},\label{eq:GDT-Susceptibility_tensor}
\end{equation}
where $\chi^{(s)}_{n,ij}\left(\boldsymbol{k},\omega\right)$ is the
harmonic contribution to the $ij$-th component of the susceptibility
tensor and $\tilde{\chi}^{(s)}_{zz}$ is an additional contribution
from the $zz$ component alone. These quantities are given by
\begin{align}
\chi^{(s)}_{n,ij}\left(\boldsymbol{k},\omega\right) & =\frac{\omega^{2}_{ps}}{\omega^{2}}\int d^{3}v\frac{v_{\perp}\left(\Pi^{*}_{ns}\right)_{i}\left(\Pi_{ns}\right)_{j}\mathcal{L}f_{s}}{\omega-n\Omega_{s}-k_{\parallel}v_{\parallel}}, & \tilde{\chi}^{(s)}_{zz} & =\frac{\omega^{2}_{ps}}{\omega^{2}}\int d^{3}v\frac{v_{\parallel}}{v_{\perp}}Lf_{s},\label{eq:GDT-ST-Harmonic}
\end{align}
where $f_{s}=f_{s}\left(v_{\parallel},v_{\perp}\right)$ is a general
VDF for species $s$, 
\begin{gather*}
\boldsymbol{\Pi}_{ns}=\frac{n}{\rho_{s}}J_{n}\left(\rho_{s}\right)\hat{\boldsymbol{x}}+iJ^{\prime}_{n}\left(\rho_{s}\right)\hat{\boldsymbol{y}}+\frac{v_{\parallel}}{v_{\perp}}J_{n}\left(\rho_{s}\right)\hat{\boldsymbol{z}},\\
\begin{aligned}Lf_{s} & =v_{\perp}\frac{\partial f_{s}}{\partial v_{\parallel}}-v_{\parallel}\frac{\partial f_{s}}{\partial v_{\perp}}, & \mathcal{L}f_{s} & =\omega\frac{\partial f_{s}}{\partial v_{\perp}}+k_{\parallel}Lf_{s},\end{aligned}
\end{gather*}
$J_{n}\left(z\right)$ is the Bessel function of the first kind \citep{OlverMaximon-Full-NIST10},
$\omega_{ps}=\sqrt{4\pi n_{s}q^{2}_{s}/m_{s}}$ and $\Omega_{s}=q_{s}B_{0}/m_{s}c$
are, respectively, the plasma and the cyclotron frequencies of particles
with charge $q_{s}$, mass $m_{s}$ and number density $n_{s}$, embedded
in a field of magnetic induction with intensity $B_{0}$, $c$ is
the vacuum light speed and $\rho_{s}=k_{\perp}v_{\perp}/\Omega_{s}$\@.
The distribution function must be normalized to one, i.e.,
$\int d^{3}v\,f_{s}\left(v_{\parallel},v_{\perp}\right)=1$ and we
have adopted the \emph{Stix frame}, i.e., $\boldsymbol{B}_{0}=B_{0}\hat{\boldsymbol{z}}$
$\left(B_{0}>0\right)$ and $\boldsymbol{k}=k_{\perp}\hat{\boldsymbol{x}}+k_{\parallel}\hat{\boldsymbol{z}}$\@.
In this frame, $\varepsilon_{yx}=-\varepsilon_{xy}$, $\varepsilon_{zx}=\varepsilon_{xz}$
and $\varepsilon_{zy}=-\varepsilon_{yz}$.

\end{subequations}

\subsubsection{The general expressions for the susceptibility tensor}\label{subsec:Susceptibility_tensor-gen}

The task we propose to endevour here is to derive for the first time
the susceptibility tensor for a bRKd plasma population, described
by the VDF (\ref{eq:RKAP3:RKD-anisotropic-beam-generalized})\@.
This task could be accomplished by the direct derivation of (\ref{eq:RKAP3:RKD-anisotropic-beam-generalized}),
followed the evaluation of the integrals in (\ref{eq:GDT-ST-Harmonic})\@.
However, we find it more convenient if we first employ the identity
\begin{equation}
\int^{\infty}_{0}d\beta\,\beta^{\zeta-1}e^{-\gamma\beta}=\Gamma\left(\zeta\right)\gamma^{-\zeta}\quad\left(\Real\, \zeta>0,\;\gamma>0\right),\label{eq:RKAP5:Superstatistics-ident}
\end{equation}
which can be easily verified from the definition of the Gamma function.
Using this identity in the suprathermal part
of $f^{(\eta,\zeta,\boldsymbol{\mu})}_{\kappa,\boldsymbol{\theta}}$,
we can write 
\begin{equation}
f^{(\eta,\zeta,\boldsymbol{\mu})}_{\kappa,\boldsymbol{\theta}}\left(v_{\parallel},v_{\perp}\right)=\frac{N^{(\eta,\zeta,\boldsymbol{\mu})}_{\kappa,\boldsymbol{\theta}}}{\Gamma\left(\zeta\right)}\int^{\infty}_{0}d\beta\,\beta^{\zeta-1}e^{-\beta}e^{-\left(v_{\parallel}-U_{\parallel}\right)^{2}/\phi^{2}_{\parallel}}e^{-v^{2}_{\perp}/\phi^{2}_{\perp}},\label{eq:RKAP5:RKD-anisotropic_beam-superstatistics}
\end{equation}
where 
\[
\boldsymbol{\phi}\left(\beta\right)=\left(\frac{\theta_{\perp}}{\sqrt{\mu_{\perp}+\beta/\eta}},\frac{\theta_{\parallel}}{\sqrt{\mu_{\parallel}+\beta/\eta}}\right).
\]

Consequently, by inserting (\ref{eq:RKAP5:RKD-anisotropic_beam-superstatistics})
into (\ref{eq:GDT-ST-Harmonic}) and defining the new integration
variables $\left\{ x,y\right\} $ as 
\begin{align}
v_{\perp} & =\phi_{\perp s}x, & v_{\parallel} & =\phi_{\parallel s}y+U_{\parallel s},\label{eq:RKAP5:Integ_variables}
\end{align}
the susceptibility components can be derived after a tedious but straightforward
algebra, during which we also define the quantities
\begin{gather*}
\begin{aligned}\tilde{\varrho}_{s}\left(\beta\right) & =\frac{\varrho_{s}}{\mu_{\perp s}+\beta/\eta_{s}}, & \varrho_{s} & =\frac{k^{2}_{\perp}\theta^{2}_{\perp s}}{2\Omega^{2}_{s}}=\frac{1}{2}\nu^{2}_{s},\\
\tilde{\xi}_{ns}\left(\beta\right) & =\frac{\omega-n\Omega_{s}}{k_{\parallel}\phi_{\parallel s}}=\xi_{ns}\sqrt{\mu_{\parallel s}+\frac{\beta}{\eta_{s}}}, & \xi_{ns} & =\frac{\omega-n\Omega_{s}}{k_{\parallel}\theta_{\parallel s}},\\
\tilde{\varsigma}_{ns}\left(\beta\right) & =\frac{\omega-n\Omega_{s}-k_{\parallel}U_{\parallel s}}{k_{\parallel}\phi_{\parallel s}}=\varsigma_{ns}\sqrt{\mu_{\parallel s}+\frac{\beta}{\eta_{s}}}, & \varsigma_{ns} & =\frac{\omega-n\Omega_{s}-k_{\parallel}U_{\parallel s}}{k_{\parallel}\theta_{\parallel s}}=\xi_{ns}-\frac{U_{\parallel s}}{\theta_{\parallel s}},
\end{aligned}
\\
\mathscr{A}^{\mu}_{\theta}\left(\beta\right)=A_{\theta s}\frac{\mu_{\parallel s}+\beta/\eta_{s}}{\mu_{\perp s}+\beta/\eta_{s}}-\frac{\mu_{\parallel s}A_{\mu s}}{\mu_{\perp s}+\beta/\eta_{s}}=A_{\theta s}\frac{\mu_{\parallel s}+\beta/\eta_{s}}{\mu_{\perp s}+\beta/\eta_{s}}+\frac{\mu_{\perp s}\mathcal{A}_{\mu s}}{\mu_{\perp s}+\beta/\eta_{s}},\\
\begin{aligned}A_{\theta s} & =1-\frac{\theta^{2}_{\perp s}}{\theta^{2}_{\parallel s}}, & \mathcal{A}_{\mu s} & =1-\frac{\mu_{\parallel s}}{\mu_{\perp s}}, & A_{\mu s} & =1-\frac{\mu_{\perp s}}{\mu_{\parallel s}}.\end{aligned}
\end{gather*}
Quantitites $A_{\theta}$, $\mathcal{A}_{\mu}$ and $A_{\mu}$ are
anisotropy parameters that quantify the different ways the distribution
$f^{(\eta,\zeta,\boldsymbol{\mu})}_{\kappa,\boldsymbol{\theta}}\left(v_{\parallel},v_{\perp}\right)$
departs from isotropy.

With the $xz$, $yz$ and $zz$ components, additional manipulations
are performed. For the $xz$ and, $yz$ components, we complete the
denominator of the integral 
\begin{align*}
I_{y} & =\int^{\infty}_{-\infty}dy\,\frac{\left(y+\frac{U_{\parallel s}}{\phi_{\parallel s}}\right)}{y-\tilde{\varsigma}_{ns}}\left(\tilde{\varsigma}_{0s}-\mathscr{A}^{\mu}_{\theta}y\right)e^{-y^{2}}\\
 & =-\mathscr{A}^{\mu}_{\theta}\int^{\infty}_{-\infty}dy\,\left(y+\frac{U_{\parallel s}}{\phi_{\parallel s}}\right)e^{-y^{2}}+\left(\tilde{\varsigma}_{0s}-\mathscr{A}^{\mu}_{\theta}\tilde{\varsigma}_{ns}\right)\int^{\infty}_{-\infty}dy\,\frac{\left(y+\frac{U_{\parallel s}}{\phi_{\parallel s}}\right)}{y-\tilde{\varsigma}_{ns}}e^{-y^{2}}.
\end{align*}
Inserting $I_{y}$ into the corresponding susceptibilities in (\ref{eq:GDT-Susceptibility_tensor}),
we have the following terms:
\begin{align*}
T_{xz} & =-\mathscr{A}^{\mu}_{\theta}\int^{\infty}_{0}dx\,x\left[\sum^{\infty}_{n\to-\infty}\frac{n}{\nu_{s}}J^{2}_{n}\left(\tilde{\nu}_{s}x\right)\right]e^{-x^{2}}\int^{\infty}_{-\infty}dy\,\left(y+\frac{U_{\parallel s}}{\phi_{\parallel s}}\right)e^{-y^{2}}=0\\
T_{yz} & =-\mathscr{A}^{\mu}_{\theta}\int^{\infty}_{0}dx\,x^{2}\left[\sum^{\infty}_{n\to-\infty}J_{n}\left(\tilde{\nu}_{s}x\right)J^{\prime}_{n}\left(\tilde{\nu}_{s}x\right)\right]e^{-x^{2}}\int^{\infty}_{-\infty}dy\,\left(y+\frac{U_{\parallel s}}{\phi_{\parallel s}}\right)e^{-y^{2}}=0.
\end{align*}
The terms between brackets are identically zero due to the identities
in Appendix \ref{sec:Sums-of-Bessel}.

For the $zz$ component, again the denominator is completed and, noticing
that 
\[
\left(\frac{\phi^{2}_{\perp s}}{\phi^{2}_{\parallel s}}\tilde{\xi}_{0s}+\mathscr{A}^{\mu}_{\theta}\frac{n\Omega_{s}}{k_{\parallel}\phi_{\parallel s}}\right)\tilde{\varsigma}_{ns}+\frac{n\Omega_{s}}{k_{\parallel}\phi_{\parallel s}}\frac{U_{\parallel s}}{\phi_{\parallel s}}=\left(\tilde{\xi}_{0s}-\mathscr{A}^{\mu}_{\theta}\tilde{\xi}_{ns}\right)\tilde{\varsigma}_{ns}+\frac{n\Omega_{s}}{k_{\parallel}\phi_{\parallel s}}\frac{U_{\parallel s}}{\phi_{\parallel s}},
\]
the following term results in $\chi^{(s)}_{zz}$:
\begin{align*}
T_{zz} & =\int^{\infty}_{0}dx\,xe^{-x^{2}}\left[\frac{\phi^{2}_{\perp s}}{\phi^{2}_{\parallel s}}\tilde{\xi}_{0s}\left(\sum^{\infty}_{n\to-\infty}J^{2}_{n}\left(\tilde{\nu}_{s}x\right)\right)+\mathscr{A}^{\mu}_{\theta}\left(\sum^{\infty}_{n\to-\infty}\frac{n\Omega_{s}}{k_{\parallel}\phi_{\parallel s}}J^{2}_{n}\left(\tilde{\nu}_{s}x\right)\right)\right]\\
 & =\frac{1}{2}\frac{\theta^{2}_{\perp s}}{\theta^{2}_{\parallel s}}\xi_{0s}\frac{\left(\mu_{\parallel s}+\beta/\eta_{s}\right)^{3/2}}{\mu_{\perp s}+\beta/\eta_{s}},
\end{align*}
where again identities in Appendix \ref{sec:Sums-of-Bessel} were
used.

\begin{subequations}
\label{eq:RKAP5:Susceptiblity_tensor-bRKd-3}

With these manipulations and further lengthy algebra, we can finally
write the susceptibility tensor of a bRKd plasma as 
\begin{equation}
\chi^{\text{(bRKd)}}_{ij}=\sum^{\infty}_{n\to-\infty}\chi^{\text{(bRKd)}}_{n,ij}+\delta_{iz}\delta_{jz}\hat{\chi}^{\text{(bRKd)}}_{zz},
\end{equation}
where
\begin{align}
\chi^{\text{(bRKd)}}_{n,xx} & =\frac{n^{2}}{\varrho_{s}}\frac{\omega^{2}_{ps}}{\omega^{2}}\left[\varsigma_{0s}\mathscr{Z}^{\left(\eta_{s},\zeta_{s},\boldsymbol{\mu}_{s}\right)}_{0,0,n}\left(\varrho_{s},\varsigma_{ns}\right)+\frac{1}{2}\partial_{\varsigma_{ns}}\mathscr{X}^{\left(\eta_{s},\zeta_{s},\boldsymbol{\mu}_{s}\right)}_{1,0,n}\left(\varrho_{s},\varsigma_{ns}\right)\right]\\
\chi^{\text{(bRKd)}}_{n,xy} & =in\frac{\omega^{2}_{ps}}{\omega^{2}}\left[\varsigma_{0s}\partial_{\varrho_{s}}\mathscr{Z}^{\left(\eta_{s},\zeta_{s},\boldsymbol{\mu}_{s}\right)}_{0,0,n}\left(\varrho_{s},\varsigma_{ns}\right)+\frac{1}{2}\partial^{2}_{\varrho_{s},\varsigma_{ns}}\mathscr{X}^{\left(\eta_{s},\zeta_{s},\boldsymbol{\mu}_{s}\right)}_{1,0,n}\left(\varrho_{s},\varsigma_{ns}\right)\right]\\
\chi^{\text{(bRKd)}}_{n,xz} & =-\frac{\theta_{\parallel s}}{\theta_{\perp s}}\frac{\omega^{2}_{ps}}{\omega^{2}}\frac{n}{\nu_{s}}\left[\varsigma_{0s}\left(\partial_{\varsigma_{ns}}\mathscr{Z}^{\left(\eta_{s},\zeta_{s},\boldsymbol{\mu}_{s}\right)}_{1,0,n}\left(\varrho_{s},\varsigma_{ns}\right)-2\frac{U_{\parallel s}}{\theta_{\parallel s}}\mathscr{Z}^{\left(\eta_{s},\zeta_{s},\boldsymbol{\mu}_{s}\right)}_{0,0,n}\left(\varrho_{s},\varsigma_{ns}\right)\right)\right.\nonumber \\
 & \left.-\varsigma_{ns}\left(\partial_{\varsigma_{ns}}\mathscr{X}^{\left(\eta_{s},\zeta_{s},\boldsymbol{\mu}_{s}\right)}_{1,0,n}\left(\varrho_{s},\varsigma_{ns}\right)-2\frac{U_{\parallel s}}{\theta_{\parallel s}}\mathscr{X}^{\left(\eta_{s},\zeta_{s},\boldsymbol{\mu}_{s}\right)}_{0,0,n}\left(\varrho_{s},\varsigma_{ns}\right)\right)\right]\\
\chi^{\text{(bRKd)}}_{n,yy} & =\frac{\omega^{2}_{ps}}{\omega^{2}}\left\{ \varsigma_{0s}\left[\frac{n^{2}}{\varrho_{s}}\mathscr{Z}^{\left(\eta_{s},\zeta_{s},\boldsymbol{\mu}_{s}\right)}_{0,0,n}\left(\varrho_{s},\varsigma_{ns}\right)-2\varrho_{s}\partial_{\varrho_{s}}\mathscr{Z}^{\left(\eta_{s},\zeta_{s},\boldsymbol{\mu}_{s}\right)}_{0,1,n}\left(\varrho_{s},\varsigma_{ns}\right)\right]\right.\nonumber \\
 & \left.+\frac{1}{2}\left[\frac{n^{2}}{\varrho_{s}}\partial_{\varsigma_{ns}}\mathscr{X}^{\left(\eta_{s},\zeta_{s},\boldsymbol{\mu}_{s}\right)}_{1,0,n}\left(\varrho_{s},\varsigma_{ns}\right)-2\varrho_{s}\partial^{2}_{\varrho_{s},\varsigma_{ns}}\mathscr{X}^{\left(\eta_{s},\zeta_{s},\boldsymbol{\mu}_{s}\right)}_{1,1,n}\left(\varrho_{s},\varsigma_{ns}\right)\right]\right\} \\
\chi^{\text{(bRKd)}}_{n,yz} & =\frac{i}{2}\frac{\theta_{\parallel s}}{\theta_{\perp s}}\frac{\omega^{2}_{ps}}{\omega^{2}}\nu_{s}\left[\varsigma_{0s}\left(\partial^{2}_{\varrho_{s},\varsigma_{ns}}\mathscr{Z}^{\left(\eta_{s},\zeta_{s},\boldsymbol{\mu}_{s}\right)}_{1,0,n}\left(\varrho_{s},\varsigma_{ns}\right)-2\frac{U_{\parallel s}}{\theta_{\parallel s}}\partial_{\varrho_{s}}\mathscr{Z}^{\left(\eta_{s},\zeta_{s},\boldsymbol{\mu}_{s}\right)}_{0,0,n}\left(\varrho_{s},\varsigma_{ns}\right)\right)\right.\nonumber \\
 & \left.-\varsigma_{ns}\left(\partial^{2}_{\varrho_{s},\varsigma_{ns}}\mathscr{X}^{\left(\eta_{s},\zeta_{s},\boldsymbol{\mu}_{s}\right)}_{1,0,n}\left(\varrho_{s},\varsigma_{ns}\right)-2\frac{U_{\parallel s}}{\theta_{\parallel s}}\partial_{\varrho_{s}}\mathscr{X}^{\left(\eta_{s},\zeta_{s},\boldsymbol{\mu}_{s}\right)}_{0,0,n}\left(\varrho_{s},\varsigma_{ns}\right)\right)\right]\\
\chi^{\text{(bRKd)}}_{n,zz} & =-\frac{\theta^{2}_{\parallel s}}{\theta^{2}_{\perp s}}\frac{\omega^{2}_{ps}}{\omega^{2}}\xi_{ns}\left[\varsigma_{0s}\left(\partial_{\varsigma_{ns}}\mathscr{Z}^{\left(\eta_{s},\zeta_{s},\boldsymbol{\mu}_{s}\right)}_{1,0,n}\left(\varrho_{s},\varsigma_{ns}\right)-2\frac{U_{\parallel s}}{\theta_{\parallel s}}\mathscr{Z}^{\left(\eta_{s},\zeta_{s},\boldsymbol{\mu}_{s}\right)}_{0,0,n}\left(\varrho_{s},\varsigma_{ns}\right)\right)\right.\nonumber \\
 & \left.-\varsigma_{ns}\left(\partial_{\varsigma_{ns}}\mathscr{X}^{\left(\eta_{s},\zeta_{s},\boldsymbol{\mu}_{s}\right)}_{1,0,n}\left(\varrho_{s},\varsigma_{ns}\right)-2\frac{U_{\parallel s}}{\theta_{\parallel s}}\mathscr{X}^{\left(\eta_{s},\zeta_{s},\boldsymbol{\mu}_{s}\right)}_{0,0,n}\left(\varrho_{s},\varsigma_{ns}\right)\right)\right]\\
\hat{\chi}^{\text{(bRKd)}}_{zz} & =\frac{2}{\eta_{s}}\frac{\mathcal{U}^{\left(\eta_{s},\zeta_{s},\boldsymbol{\mu}_{s}\right)}_{-1,0}}{\mathcal{U}^{\left(\eta_{s},\zeta_{s},\boldsymbol{\mu}_{s}\right)}}\frac{\omega^{2}_{ps}}{\omega^{2}}\frac{U_{\parallel s}}{\theta_{\parallel s}}\xi_{0s}.
\end{align}
All components are written in terms of the two-variables special functions
$\mathscr{Z}^{\left(\eta,\zeta,\boldsymbol{\mu}\right)}_{p_{1},p_{2},n}\left(\varrho,\varsigma\right)$
and $\mathscr{X}^{\left(\eta,\zeta,\boldsymbol{\mu}\right)}_{p_{1},p_{2},n}\left(\varrho,\varsigma\right)$
that are defined and discussed in section \ref{sec:The-special-functions},
as well as their partial derivatives, where, for instance, 
\begin{align*}
\partial_{\varrho}\mathscr{Z}^{\left(\eta,\zeta,\boldsymbol{\mu}\right)}_{p_{1},p_{2},n} & =\frac{\partial\hphantom{\varrho}}{\partial\varrho}\mathscr{Z}^{\left(\eta,\zeta,\boldsymbol{\mu}\right)}_{p_{1},p_{2},n}, & \partial_{\varsigma}\mathscr{Z}^{\left(\eta,\zeta,\boldsymbol{\mu}\right)}_{p_{1},p_{2},n} & =\frac{\partial\hphantom{\varsigma}}{\partial\varsigma}\mathscr{Z}^{\left(\eta,\zeta,\boldsymbol{\mu}\right)}_{p_{1},p_{2},n}, & \partial^{2}_{\varrho,\varsigma}\mathscr{Z}^{\left(\eta,\zeta,\boldsymbol{\mu}\right)}_{p_{1},p_{2},n} & =\frac{\partial^{2}\hphantom{\partial\varsigma}}{\partial\varrho\partial\varsigma}\mathscr{Z}^{\left(\eta,\zeta,\boldsymbol{\mu}\right)}_{p_{1},p_{2},n}.
\end{align*}
\end{subequations}
The constants $\mathcal{U}^{\left(\eta_{s},\zeta_{s},\boldsymbol{\mu}_{s}\right)}_{-1,0}$ and $\mathcal{U}^{\left(\eta_{s},\zeta_{s},\boldsymbol{\mu}_{s}\right)}$ are defined in section \ref{sec:Suprathermal_function_constant}.

\subsubsection{Particular cases of the susceptibility tensor}

Several important particular cases of the general tensor (\ref{eq:RKAP5:Susceptiblity_tensor-bRKd-3})
can be presently derived.

\paragraph{\textbf{Limit of parallel propagation}}

The first particular case we can consider is the case of parallel
propagation of waves $\left(k_{\perp}=0\right)$, given by the limit
$\varrho=0$ of functions $\mathscr{Z}^{\left(\eta,\zeta,\boldsymbol{\mu}\right)}_{p_{1},p_{2},n}\left(\varrho,\varsigma\right)$
and $\mathscr{X}^{\left(\eta,\zeta,\boldsymbol{\mu}\right)}_{p_{1},p_{2},n}\left(\varrho,\varsigma\right)$\@.
These limits are given by (\ref{eq:RKAP5:ZX-rho0}), in which case
the susceptibility tensor components in (\ref{eq:RKAP5:Susceptiblity_tensor-bRKd-3})
reduce to $\chi_{s,xz}=\chi_{s,yz}=0$, $\chi_{s,xx}=\chi_{s,yy}=\chi_{s,1}$,
$\chi_{s,xy}=\chi_{s,2}$, and $\chi_{s,zz}=\chi_{s,3}$, where 
\begin{align*}
\chi^{\text{(bRKd)}}_{s,1} & =\frac{1}{2}\frac{\omega^{2}_{ps}}{\omega^{2}}\sum_{n=\pm1}\left[\varsigma_{0s}Z^{\left(\eta_{s},\zeta_{s},\boldsymbol{\mu}_{s}\right)}_{\kappa}\left(\varsigma_{ns}\right)+\frac{1}{2}\mathcal{Z}^{\left(\eta_{s},\zeta_{s},\boldsymbol{\mu}_{s}\right)\prime}_{\kappa,\boldsymbol{\theta}_{s}}\left(\varsigma_{ns}\right)\right]\\
\chi^{\text{(bRKd)}}_{s,2} & =\frac{i}{2}\frac{\omega^{2}_{ps}}{\omega^{2}}\sum_{n=\pm1}n\left[\varsigma_{0s}Z^{\left(\eta_{s},\zeta_{s},\boldsymbol{\mu}_{s}\right)}_{\kappa}\left(\varsigma_{ns}\right)+\frac{1}{2}\mathcal{Z}^{\left(\eta_{s},\zeta_{s},\boldsymbol{\mu}_{s}\right)\prime}_{\kappa,\boldsymbol{\theta}_{s}}\left(\varsigma_{ns}\right)\right]\\
\chi^{\text{(bRKd)}}_{s,3} & =-\frac{\omega^{2}_{ps}}{\omega^{2}}\xi^{2}_{0s}Z^{\left(\eta_{s},\zeta_{s},\boldsymbol{\mu}_{s}\right)\prime}_{\kappa}\left(\varsigma_{0s}\right),
\end{align*}
written in terms of the plasma dispersion functions (\ref{eq:RKAP5:ZRKDFs-Parallel}),
for the case of parallel-propagating waves. The final expression for
$\chi^{\text{(bRKd)}}_{s,3}$ also involved identities (\ref{eq:RKAP5:ARKDF-Z-Z_cal} and \ref{eq:RKAP5:dZ-dsig-RR})\@.
This result completely agrees with the susceptibility components derived
by \citet{Gaelzer+26/06-arxiv}.

\paragraph{\textbf{Isotropic regularization parameter}}

\begin{subequations}
\label{eq:RKAP5:Susceptiblity_tensor-bRKd-Mu_isotropic-2}

Another important particular case is $\mu_{\perp}=\mu_{\parallel}=\mu$\@.
In this case, $A_{\mu}=\mathcal{A}_{\mu}=0$ and the only remaining
anisotropy is on the temperatures $\left(A_{\theta}\neq0\right)$\@.
Consequently, from (\ref{eq:RKAP5:ZXD-X-Z_relation}), 
\[
\mathscr{X}^{\left(\eta,\zeta,\mu\right)}_{p_{1},p_{2},n}\left(\varrho,\varsigma\right)=A_{\theta}\mathscr{Z}^{\left(\eta,\zeta,\mu\right)}_{p_{1}-1,p_{2}+1,n}\left(\varrho,\varsigma\right),
\]
where $\mathscr{Z}^{\left(\eta,\zeta,\mu\right)}_{p_{1},p_{2},n}\left(\varrho,\varsigma\right)$
is the limit form (\ref{eq:Mu-isotropic_limit})\@. These particular
forms are then directly inserted in (\ref{eq:RKAP5:Susceptiblity_tensor-bRKd-3}),
which will then reduce to the $\mu$-isotropic limit of the susceptibility
tensor,
\begin{align}
\chi^{\text{(bRKd)}}_{ns,xx} & =\frac{n^{2}}{\varrho_{s}}\frac{\omega^{2}_{ps}}{\omega^{2}}\left[\varsigma_{0s}\mathscr{Z}^{\left(\eta_{s},\zeta_{s},\mu_{s}\right)}_{0,0,n}\left(\varrho_{s},\varsigma_{ns}\right)+\frac{1}{2}A_{\theta s}\partial_{\varsigma_{ns}}\mathscr{Z}^{\left(\eta_{s},\zeta_{s},\mu_{s}\right)}_{0,1,n}\left(\varrho_{s},\varsigma_{ns}\right)\right]\\
\chi^{\text{(bRKd)}}_{ns,xy} & =in\frac{\omega^{2}_{ps}}{\omega^{2}}\left[\varsigma_{0s}\partial_{\varrho_{s}}\mathscr{Z}^{\left(\eta_{s},\zeta_{s},\mu_{s}\right)}_{0,0,n}\left(\varrho_{s},\varsigma_{ns}\right)+\frac{1}{2}A_{\theta s}\partial^{2}_{\varrho_{s},\varsigma_{ns}}\mathscr{Z}^{\left(\eta_{s},\zeta_{s},\mu_{s}\right)}_{0,1,n}\left(\varrho_{s},\varsigma_{ns}\right)\right]\\
\chi^{\text{(bRKd)}}_{ns,xz} & =-\frac{\theta_{\parallel s}}{\theta_{\perp s}}\frac{\omega^{2}_{ps}}{\omega^{2}}\frac{n}{\nu_{s}}\left(\varsigma_{0s}-A_{\theta s}\varsigma_{ns}\right)\left[
\vphantom{\frac{U_{\parallel s}}{\theta_{\parallel s}}}
\partial_{\varsigma_{ns}}\mathscr{Z}^{\left(\eta_{s},\zeta_{s},\mu_{s}\right)}_{1,0,n}\left(\varrho_{s},\varsigma_{ns}\right) \right.\nonumber \\
& \left. -2\frac{U_{\parallel s}}{\theta_{\parallel s}}\mathscr{Z}^{\left(\eta_{s},\zeta_{s},\mu_{s}\right)}_{0,0,n}\left(\varrho_{s},\varsigma_{ns}\right)\right]\\
\chi^{\text{(bRKd)}}_{ns,yy} & =\frac{\omega^{2}_{ps}}{\omega^{2}}\left\{ \varsigma_{0s}\left[\frac{n^{2}}{\varrho_{s}}\mathscr{Z}^{\left(\eta_{s},\zeta_{s},\mu_{s}\right)}_{0,0,n}\left(\varrho_{s},\varsigma_{ns}\right)-2\varrho_{s}\partial_{\varrho_{s}}\mathscr{Z}^{\left(\eta_{s},\zeta_{s},\mu_{s}\right)}_{0,1,n}\left(\varrho_{s},\varsigma_{ns}\right)\right]\right.\nonumber \\
 & \left.+\frac{1}{2}A_{\theta s}\left[\frac{n^{2}}{\varrho_{s}}\partial_{\varsigma_{ns}}\mathscr{Z}^{\left(\eta_{s},\zeta_{s},\mu_{s}\right)}_{0,1,n}\left(\varrho_{s},\varsigma_{ns}\right)-2\varrho_{s}\partial^{2}_{\varrho_{s},\varsigma_{ns}}\mathscr{Z}^{\left(\eta_{s},\zeta_{s},\mu_{s}\right)}_{0,2,n}\left(\varrho_{s},\varsigma_{ns}\right)\right]\right\} \\
\chi^{\text{(bRKd)}}_{ns,yz} & =\frac{i}{2}\frac{\theta_{\parallel s}}{\theta_{\perp s}}\frac{\omega^{2}_{ps}}{\omega^{2}}\nu_{s}\left(\varsigma_{0s}-A_{\theta s}\varsigma_{ns}\right)\left[
\vphantom{\frac{U_{\parallel s}}{\theta_{\parallel s}}}
\partial^{2}_{\varrho_{s},\varsigma_{ns}}\mathscr{Z}^{\left(\eta_{s},\zeta_{s},\mu_{s}\right)}_{1,0,n}\left(\varrho_{s},\varsigma_{ns}\right)\right.\nonumber \\
 & \left.-2\frac{U_{\parallel s}}{\theta_{\parallel s}}\partial_{\varrho_{s}}\mathscr{Z}^{\left(\eta_{s},\zeta_{s},\mu_{s}\right)}_{0,0,n}\left(\varrho_{s},\varsigma_{ns}\right)\right]\\
\chi^{\text{(bRKd)}}_{ns,zz} & =-\frac{\theta^{2}_{\parallel s}}{\theta^{2}_{\perp s}}\frac{\omega^{2}_{ps}}{\omega^{2}}\xi_{ns}\left(\varsigma_{0s}-A_{\theta s}\varsigma_{ns}\right)\left[
\vphantom{\frac{U_{\parallel s}}{\theta_{\parallel s}}}
\partial_{\varsigma_{ns}}\mathscr{Z}^{\left(\eta_{s},\zeta_{s},\mu_{s}\right)}_{1,0,n}\left(\varrho_{s},\varsigma_{ns}\right)\right.\nonumber \\
 & \left.-2\frac{U_{\parallel s}}{\theta_{\parallel s}}\mathscr{Z}^{\left(\eta_{s},\zeta_{s},\mu_{s}\right)}_{0,0,n}\left(\varrho_{s},\varsigma_{ns}\right)\right]\\
\hat{\chi}^{\text{(bRKd)}}_{s,zz} & =\frac{2}{\eta_{s}}\frac{\mathcal{U}^{\left(\eta_{s},\zeta_{s},\mu_{s}\right)}_{-1,0}}{\mathcal{U}^{\left(\eta_{s},\zeta_{s},\mu_{s}\right)}}\frac{\omega^{2}_{ps}}{\omega^{2}}\frac{U_{\parallel s}}{\theta_{\parallel s}}\xi_{0s}.
\end{align}
\end{subequations}

\paragraph{\textbf{Standard Kappa limit}}

The final important case is the limit $\mu=0$, in which case (\ref{eq:RKAP5:Susceptiblity_tensor-bRKd-Mu_isotropic-2})
will reduce to the susceptibility tensor of a bi-standard Kappa distribution
with drift (bSKd)\@. In this case, $\mathscr{X}^{\left(\eta,\zeta,0\right)}_{p_{1},p_{2},n}\left(\varrho,\varsigma\right)=A_{\theta}\mathscr{Z}^{\left(\eta,\zeta,0\right)}_{p_{1}-1,p_{2}+1,n}\left(\varrho,\varsigma\right)$,
which is then given by (\ref{eq:SKDL-1})\@.

\begin{subequations}
\label{eq:RKAP5:bSKd_limit}

However, at this point it is more interesting to rewrite the tensor
components according to expressions previously published in the literature,
namely, in terms of the special function $\mathcal{Z}^{\left(\alpha,\beta\right)}_{n,\kappa}\left(\varrho,\varsigma\right)$
discussed by \citet{GaelzerZiebell16/02,Gaelzer+16/06}\@. This change
of notation will be called the RKD $\longrightarrow$ SKD transition
and it can be accomplished by the recipe $\eta\to\kappa$, $\zeta\to\sigma=\kappa+\alpha$
and $p_{1}+p_{2}\to2-\beta$\@. Proceeding according to this recipe,
we identify (\ref{eq:SKDL-2}) and obtain the susceptibility tensor
for a bSKd plasma, 
\begin{align}
\chi^{\text{\text{\text{(bSKd)}}}}_{n,xx} & =\frac{n^{2}}{\varrho_{s}}\frac{\omega^{2}_{ps}}{\omega^{2}}\left[\varsigma_{0s}\mathcal{Z}^{\left(\alpha_{s},2\right)}_{n,\kappa_{s}}\left(\varrho_{s},\varsigma_{ns}\right)+\frac{1}{2}A_{\theta s}\partial_{\varsigma_{ns}}\mathcal{Z}^{\left(\alpha_{s},1\right)}_{n,\kappa_{s}}\left(\varrho_{s},\varsigma_{ns}\right)\right]\\
\chi^{\text{\text{\text{\text{(bSKd)}}}}}_{n,xy} & =in\frac{\omega^{2}_{ps}}{\omega^{2}}\left[\varsigma_{0s}\partial_{\varrho_{s}}\mathcal{Z}^{\left(\alpha_{s},2\right)}_{n,\kappa_{s}}\left(\varrho_{s},\varsigma_{ns}\right)+\frac{1}{2}A_{\theta s}\partial^{2}_{\varrho_{s},\varsigma_{ns}}\mathcal{Z}^{\left(\alpha_{s},1\right)}_{n,\kappa_{s}}\left(\varrho_{s},\varsigma_{ns}\right)\right]\\
\chi^{\text{\text{\text{\text{(bSKd)}}}}}_{n,xz} & =-\frac{\theta_{\parallel s}}{\theta_{\perp s}}\frac{\omega^{2}_{ps}}{\omega^{2}}\frac{n}{\nu_{s}}\left(\varsigma_{0s}-A_{\theta s}\varsigma_{ns}\right)\left[\partial_{\varsigma_{ns}}\mathcal{Z}^{\left(\alpha_{s},1\right)}_{n,\kappa_{s}}\left(\varrho_{s},\varsigma_{ns}\right)-2\frac{U_{\parallel s}}{\theta_{\parallel s}}\mathcal{Z}^{\left(\alpha_{s},2\right)}_{n,\kappa_{s}}\left(\varrho_{s},\varsigma_{ns}\right)\right]\\
\chi^{\text{\text{\text{\text{(bSKd)}}}}}_{n,yy} & =\frac{\omega^{2}_{ps}}{\omega^{2}}\left\{ \varsigma_{0s}\left[\frac{n^{2}}{\varrho_{s}}\mathcal{Z}^{\left(\alpha_{s},2\right)}_{n,\kappa_{s}}\left(\varrho_{s},\varsigma_{ns}\right)-2\varrho_{s}\partial_{\varrho_{s}}\mathcal{Z}^{\left(\alpha_{s},1\right)}_{n,\kappa_{s}}\left(\varrho_{s},\varsigma_{ns}\right)\right]\right.\nonumber \\
 & \left.+\frac{1}{2}A_{\theta s}\left[\frac{n^{2}}{\varrho_{s}}\partial_{\varsigma_{ns}}\mathcal{Z}^{\left(\alpha_{s},1\right)}_{n,\kappa_{s}}\left(\varrho_{s},\varsigma_{ns}\right)-2\varrho_{s}\partial^{2}_{\varrho_{s},\varsigma_{ns}}\mathcal{Z}^{\left(\alpha_{s},0\right)}_{n,\kappa_{s}}\left(\varrho_{s},\varsigma_{ns}\right)\right]\right\} \\
\chi^{\text{\text{\text{\text{(bSKd)}}}}}_{n,yz} & =\frac{i}{2}\frac{\theta_{\parallel s}}{\theta_{\perp s}}\frac{\omega^{2}_{ps}}{\omega^{2}}\nu_{s}\left(\varsigma_{0s}-A_{\theta s}\varsigma_{ns}\right)\left[\partial^{2}_{\varrho_{s},\varsigma_{ns}}\mathcal{Z}^{\left(\alpha_{s},1\right)}_{n,\kappa_{s}}\left(\varrho_{s},\varsigma_{ns}\right)-2\frac{U_{\parallel s}}{\theta_{\parallel s}}\partial_{\varrho_{s}}\mathcal{Z}^{\left(\alpha_{s},2\right)}_{n,\kappa_{s}}\left(\varrho_{s},\varsigma_{ns}\right)\right]\\
\chi^{\text{\text{\text{\text{(bSKd)}}}}}_{n,zz} & =-\frac{\theta^{2}_{\parallel s}}{\theta^{2}_{\perp s}}\frac{\omega^{2}_{ps}}{\omega^{2}}\xi_{ns}\left(\varsigma_{0s}-A_{\theta s}\varsigma_{ns}\right)\left[\partial_{\varsigma_{ns}}\mathcal{Z}^{\left(\alpha_{s},1\right)}_{n,\kappa_{s}}\left(\varrho_{s},\varsigma_{ns}\right)-2\frac{U_{\parallel s}}{\theta_{\parallel s}}\mathcal{Z}^{\left(\alpha_{s},2\right)}_{n,\kappa_{s}}\left(\varrho_{s},\varsigma_{ns}\right)\right]\\
\hat{\chi}^{\text{\text{\text{\text{(bSKd)}}}}}_{zz} & =2\frac{\sigma_{s}-\nicefrac{3}{2}}{\kappa_{s}}\frac{\omega^{2}_{ps}}{\omega^{2}}\frac{U_{\parallel s}}{\theta_{\parallel s}}\xi_{0s},
\end{align}
where we have also used (\ref{eq:RKAP5:U_cal-fun-Mu0}) at $z=0$.
\end{subequations}

Tensor $\chi^{\text{\text{\text{(bSKd)}}}}_{n,ij}$ generalizes the
expressions given by \citet{Gaelzer+16/06}, which correspond to the
particular case $U_{\parallel s}=0$ of (\ref{eq:RKAP5:bSKd_limit})\@.
Another important point to mention is that in \citet{Gaelzer+16/06},
the $\chi^{\text{\text{\text{(bSKd)}}}}_{n,yy}$ component was written
in terms of the function $\mathcal{Y}^{\left(\alpha,\beta\right)}_{n,\kappa}\left(\varrho,\varsigma\right)$,
but subsequently we realized that 
\begin{equation}
\mathcal{Y}^{\left(\alpha,\beta\right)}_{n,\kappa}\left(\varrho,\varsigma\right)=\partial_{\varrho}\mathcal{Z}^{\left(\alpha,\beta-1\right)}_{n,\kappa}\left(\varrho,\varsigma\right).\label{eq:KAP7:KPSF2V-Y-dZ}
\end{equation}
The proof of this result is given in Appendix \ref{sec:Proof-of-eq_9}.

\paragraph{\textbf{Maxwellian limit}}

The final case is the Maxwellian limit, obtained by performing $\kappa\to\infty$
in (\ref{eq:RKAP5:bSKd_limit})\@. In this case the function $\mathcal{Z}^{\left(\alpha,\beta\right)}_{n,\kappa}\left(\varrho,\varsigma\right)$
factors according to (\ref{eq:Maxwellian_limit}) and we readily obtain
the susceptibility tensor for a bi-Maxwellian with drift distribution.

\section{The special functions for the susceptibility tensor of a bRKd plasma}\label{sec:The-special-functions}

More often than not, new special functions need to be defined and
analysed whenever someone endeavours to obtain an analytical description
of the behaviour of wave propagation and wave-particle interaction
in a plasma whose statistical properties are modelled by a new velocity
distribution function. Those new functions are needed if one wants
to relate the new susceptibility tensor components with previously
known expressions and also if one wants to derive approximations that
can lead to simple analytical expressions for the dispersion relations
of the wave normal modes and for the damping/growth rate coefficients.

In this section, the new special functions that sprout from the susceptibility
tensor (\ref{eq:GDT-Susceptibility_tensor}) for a bRKd plasma are
defined and analysed. Basically, three two-variables functions come
forth from the analysis, with two one-variable functions that follow
as particular cases of the former and an additional pair of one-variable
function and constant. These functions generalize results from previous
publications 
\citep{GaelzerZiebell14/12,GaelzerZiebell16/02,Gaelzer+16/06,Gaelzer+24/07,%
Gaelzer+26/06-arxiv}\@.
Alternative and equivalent representations for the functions will
be presented and basic mathematical properties are derived, that can
provide immediate numerical implementations.

\subsection{The two-variables functions $\mathscr{Z}^{\left(\eta,\zeta,\boldsymbol{\mu}\right)}_{p_{1},p_{2},n}\left(\varrho,\varsigma\right)$
and $\mathscr{X}^{\left(\eta,\zeta,\boldsymbol{\mu}\right)}_{p_{1},p_{2},n}\left(\varrho,\varsigma\right)$}

Here, the new functions $\mathscr{Z}^{\left(\eta,\zeta,\boldsymbol{\mu}\right)}_{p_{1},p_{2},n}\left(\varrho,\varsigma\right)$
and $\mathscr{X}^{\left(\eta,\zeta,\boldsymbol{\mu}\right)}_{p_{1},p_{2},n}\left(\varrho,\varsigma\right)$
that appear in all components of the susceptibility tensor (\ref{eq:RKAP5:Susceptiblity_tensor-bRKd-3})
are defined through different representations and some of their properties
are discussed. The first argument, $\varrho$, is related to the quantity
$k_{\perp}r_{L}$, i.e., the product of the particle's gyroradius
$r_{L}$ (or Larmor radius) and the perpendicular wave vector component.
Hence, this argument is relevant to analyse oblique and perpendicular
wave propagation, relative to $\boldsymbol{B}_{0}$\@. On the other
hand, the second argument, $\varsigma$, is related to $k_{\parallel}\theta_{\parallel}/\omega$,
i.e., the ratio of the particle's parallel thermal speed with
the parallel component of the wave's phase velocity, including also
the effects of Doppler shifts due to the harmonics of the cyclotron
frequency and the drift speeds.

More than a decade of efforts with the special functions that describe
wave-particle interactions in suprathermal plasmas have been unsuccessful
to factor out the two-variables functions as a simple product of one-variable
functions, such as $\mathscr{Z}\left(\varrho,\varsigma\right)=R\left(\varrho\right)S\left(\varsigma\right)$,
as is the case with Maxwellian-like distributions. We are now convinced
that the intrinsic transcendental nature of these functions on their
arguments is a mathematical consequence of the physical meaning of
a (quasi-) steady state of a statistical system away from thermal
equilibrium. A system in thermal equilibrium is characterized by short-ranged
Coulombian collisions that result in the destruction of any correlation
among the different degrees of freedom. As a consequence, the Maxwell-Boltzmann
distribution function that statistically describes the thermal equilibrium
of such a system is exactly separable in all its degrees of freedom.
In contrast, space plasmas in a quasi-equilibrium state are usually
almost noncollisional (within a characteristic time scale) and consequently
display strong correlations among its degrees of freedom. Hence, the
intrinsic transcendental nature of the special functions that describe
wave-particle interactions in systems that are in nonthermal equilibrium
states is a mathematical consequence of their physical nature.

\subsubsection{Representations}

The first set of representations follow directly from the evaluation
of the integrals in (\ref{eq:GDT-ST-Harmonic}) with the form (\ref{eq:RKAP5:RKD-anisotropic_beam-superstatistics})
of the distribution. After defining the nondimensional integration
variables $\left\{ x,y\right\} $ according to (\ref{eq:RKAP5:Integ_variables}),
one can forthwith identify in all tensor components the well-known
Fried \& Conte function \citep{FriedConte61}
\begin{equation}
Z\left(\zeta\right)=\frac{1}{\sqrt{\pi}}\int^{\infty}_{-\infty}\frac{e^{-y^{2}}dy}{y-\zeta}\quad\left(\Imag\, \zeta>0\right)\label{eq:Fried-Conte_function}
\end{equation}
and its derivative. Additionally, from the identity \citep{GradshteynRyzhik07}
\begin{equation}
\int^{\infty}_{0}dx\,xJ_{p}\left(\alpha x\right)J_{p}\left(\beta x\right)e^{-\rho^{2}x^{2}}=\frac{1}{2\rho^{2}}\exp\left(-\frac{\alpha^{2}+\beta^{2}}{4\rho^{2}}\right)I_{p}\left(\frac{\alpha\beta}{2\rho^{2}}\right),\label{eq:GMP01:Int-JJe}
\end{equation}
where $J_{p}\left(z\right)$ is the Bessel function and $I_{p}\left(z\right)$
the modified Bessel function of the first kind \citep{OlverMaximon-Full-NIST10},
one readily identifies 
\begin{equation}
\mathscr{H}_{n}\left(z\right)=e^{-z}I_{n}\left(z\right)=\frac{z^{\left|n\right|}}{2^{\left|n\right|}\left|n\right|!}\pFq 11\left({\left|n\right|+\nicefrac{1}{2}\atop 2\left|n\right|+1};-2z\right)=\frac{z^{\left|n\right|}}{2^{\left|n\right|}\left|n\right|!}\sum^{\infty}_{k=0}\frac{\left(\left|n\right|+\nicefrac{1}{2}\right)_{k}}{\left(2\left|n\right|+1\right)_{k}}\frac{\left(-2z\right)^{k}}{k!},\label{eq:Gyroradius_function}
\end{equation}
called here the plasma gyroradius function. In (\ref{eq:Gyroradius_function}),
$\pFq 11\left(\cdots\right)$ is the Kummer confluent hypergeometric
function \citep{Daalhuis-Full-NIST10a} and $\left(a\right)_{k}$
is the Pochhammer symbol \citep{AskeyRoy-Full-NIST10}. 

\begin{subequations}
\label{eq:RKAP5:Z_scr-X_scr-Derivs}

Proceeding along these guidelines, one can define the functions
\begin{align}
\mathscr{Z}^{\left(\eta,\zeta,\boldsymbol{\mu}\right)}_{p_{1},p_{2},n}\left(\varrho,\varsigma\right) & =\mathcal{N}^{\left(\eta,\zeta,\boldsymbol{\mu}\right)}\int^{\infty}_{0}d\beta\,\frac{\beta^{\zeta-1}e^{-\beta}\mathscr{H}_{n}\left(\tilde{\varrho}\right)Z\left(\tilde{\varsigma}\right)}{\left(\mu_{\parallel}+\beta/\eta\right)^{p_{1}}\left(\mu_{\perp}+\beta/\eta\right)^{p_{2}}}\label{eq:RKAP5:ZXD-Z_scr}\\
\mathscr{X}^{\left(\eta,\zeta,\boldsymbol{\mu}\right)}_{p_{1},p_{2},n}\left(\varrho,\varsigma\right) & =\mathcal{N}^{\left(\eta,\zeta,\boldsymbol{\mu}\right)}\int^{\infty}_{0}d\beta\,\frac{\mathscr{A}^{\mu}_{\theta}\beta^{\zeta-1}e^{-\beta}\mathscr{H}_{n}\left(\tilde{\varrho}\right)Z\left(\tilde{\varsigma}\right)}{\left(\mu_{\parallel}+\beta/\eta\right)^{p_{1}}\left(\mu_{\perp}+\beta/\eta\right)^{p_{2}}},\label{eq:RKAP5:ZXD-X_scr}
\end{align}
where $\mathcal{N}^{\left(\eta,\zeta,\boldsymbol{\mu}\right)}$ is
given by (\ref{eq:Nc-gen})  and the quantities $\tilde{\varrho}\left(\beta\right)$,
$\tilde{\varsigma}\left(\beta\right)$ and $\mathscr{A}^{\mu}_{\theta}\left(\beta\right)$
are defined in section \ref{subsec:Susceptibility_tensor-gen}. 

From the definitions above, one readily evaluates the derivatives
\begin{align}
\partial_{\varsigma}\mathscr{Z}^{\left(\eta,\zeta,\boldsymbol{\mu}\right)}_{p_{1},p_{2},n}\left(\varrho,\varsigma\right) & =\mathcal{N}^{\left(\eta,\zeta,\boldsymbol{\mu}\right)}\int^{\infty}_{0}d\beta\,\frac{\beta^{\zeta-1}e^{-\beta}\mathscr{H}_{n}\left(\tilde{\varrho}\right)Z^{\prime}\left(\tilde{\varsigma}\right)}{\left(\mu_{\parallel}+\beta/\eta\right)^{p_{1}-1/2}\left(\mu_{\perp}+\beta/\eta\right)^{p_{2}}}\label{eq:RKAP5:ZXD-dZ_scr-dsigma}\\
\partial_{\varrho}\mathscr{Z}^{\left(\eta,\zeta,\boldsymbol{\mu}\right)}_{p_{1},p_{2},n}\left(\varrho,\varsigma\right) & =\mathcal{N}^{\left(\eta,\zeta,\boldsymbol{\mu}\right)}\int^{\infty}_{0}d\beta\,\frac{\beta^{\zeta-1}e^{-\beta}\mathscr{H}^{\prime}_{n}\left(\tilde{\varrho}\right)Z\left(\tilde{\varsigma}\right)}{\left(\mu_{\parallel}+\beta/\eta\right)^{p_{1}}\left(\mu_{\perp}+\beta/\eta\right)^{p_{2}+1}}\label{eq:ZXD-dZ_scr-drho}\\
\partial_{\varsigma}\partial_{\varrho}\mathscr{Z}^{\left(\eta,\zeta,\boldsymbol{\mu}\right)}_{p_{1},p_{2},n}\left(\varrho,\varsigma\right) & =\mathcal{N}^{\left(\eta,\zeta,\boldsymbol{\mu}\right)}\int^{\infty}_{0}d\beta\,\frac{\beta^{\zeta-1}e^{-\beta}\mathscr{H}^{\prime}_{n}\left(\tilde{\varrho}\right)Z^{\prime}\left(\tilde{\varsigma}\right)}{\left(\mu_{\parallel}+\beta/\eta\right)^{p_{1}-1/2}\left(\mu_{\perp}+\beta/\eta\right)^{p_{2}+1}}\\
\partial_{\varsigma}\mathscr{X}^{\left(\eta,\zeta,\boldsymbol{\mu}\right)}_{p_{1},p_{2},n}\left(\varrho,\varsigma\right) & =\mathcal{N}^{\left(\eta,\zeta,\boldsymbol{\mu}\right)}\int^{\infty}_{0}d\beta\,\frac{\mathscr{A}^{\boldsymbol{\mu}}_{\boldsymbol{\theta}}\beta^{\zeta-1}e^{-\beta}\mathscr{H}_{n}\left(\tilde{\varrho}\right)Z^{\prime}\left(\tilde{\varsigma}\right)}{\left(\mu_{\parallel}+\beta/\eta\right)^{p_{1}-1/2}\left(\mu_{\perp}+\beta/\eta\right)^{p_{2}}}\\
\partial_{\varrho}\mathscr{X}^{\left(\eta,\zeta,\boldsymbol{\mu}\right)}_{p_{1},p_{2},n}\left(\varrho,\varsigma\right) & =\mathcal{N}^{\left(\eta,\zeta,\boldsymbol{\mu}\right)}\int^{\infty}_{0}d\beta\,\frac{\mathscr{A}^{\boldsymbol{\mu}}_{\boldsymbol{\theta}}\beta^{\zeta-1}e^{-\beta}\mathscr{H}^{\prime}_{n}\left(\tilde{\varrho}\right)Z\left(\tilde{\varsigma}\right)}{\left(\mu_{\parallel}+\beta/\eta\right)^{p_{1}}\left(\mu_{\perp}+\beta/\eta\right)^{p_{2}+1}}\\
\partial_{\varsigma}\partial_{\varrho}\mathscr{X}^{\left(\eta,\zeta,\boldsymbol{\mu}\right)}_{p_{1},p_{2},n}\left(\varrho,\varsigma\right) & =\mathcal{N}^{\left(\eta,\zeta,\boldsymbol{\mu}\right)}\int^{\infty}_{0}d\beta\,\frac{\mathscr{A}^{\boldsymbol{\mu}}_{\boldsymbol{\theta}}\beta^{\zeta-1}e^{-\beta}\mathscr{H}^{\prime}_{n}\left(\tilde{\varrho}\right)Z^{\prime}\left(\tilde{\varsigma}\right)}{\left(\mu_{\parallel}+\beta/\eta\right)^{p_{1}-1/2}\left(\mu_{\perp}+\beta/\eta\right)^{p_{2}+1}},
\end{align}
where $Z^{\prime}\left(\zeta\right)=\mathrm{d}Z/\mathrm{d}\zeta$ 
and $\mathscr{H}^{\prime}_{n}\left(z\right)=\mathrm{d}\mathscr{H}_{n}/\mathrm{d}z$\@.
One can easily verify that $\mathscr{X}^{\left(\eta,\zeta,\boldsymbol{\mu}\right)}_{p_{1},p_{2},n}\left(\varrho,\varsigma\right)$
can ultimately be written in terms of $\mathscr{Z}^{\left(\eta,\zeta,\boldsymbol{\mu}\right)}_{p_{1},p_{2},n}\left(\varrho,\varsigma\right)$
in three equivalent ways:
\begin{equation}
\begin{aligned}\mathscr{X}^{\left(\eta,\zeta,\boldsymbol{\mu}\right)}_{p_{1},p_{2},n}\left(\varrho,\varsigma\right) & =\mathscr{Z}^{\left(\eta,\zeta,\boldsymbol{\mu}\right)}_{p_{1},p_{2},n}\left(\varrho,\varsigma\right)-\frac{\theta^{2}_{\perp}}{\theta^{2}_{\parallel}}\mathscr{Z}^{\left(\eta,\zeta,\boldsymbol{\mu}\right)}_{p_{1}-1,p_{2}+1,n}\left(\varrho,\varsigma\right)\\
 & =A_{\theta}\mathscr{Z}^{\left(\eta,\zeta,\boldsymbol{\mu}\right)}_{p_{1}-1,p_{2}+1,n}\left(\varrho,\varsigma\right)-\mu_{\parallel}A_{\mu}\mathscr{Z}^{\left(\eta,\zeta,\boldsymbol{\mu}\right)}_{p_{1},p_{2}+1,n}\left(\varrho,\varsigma\right)\\
 & =A_{\theta}\mathscr{Z}^{\left(\eta,\zeta,\boldsymbol{\mu}\right)}_{p_{1}-1,p_{2}+1,n}\left(\varrho,\varsigma\right)+\mu_{\perp}\mathcal{A}_{\mu}\mathscr{Z}^{\left(\eta,\zeta,\boldsymbol{\mu}\right)}_{p_{1},p_{2}+1,n}\left(\varrho,\varsigma\right),
\end{aligned}
\label{eq:RKAP5:ZXD-X-Z_relation}
\end{equation}
depending on the convenience.
\end{subequations}

Expressions (\ref{eq:RKAP5:Z_scr-X_scr-Derivs}) already provide \emph{computable
representations }for the functions, meaning that they can be readily
implemented as computer software code, since for all major programming
languages there are libraries that provide procedures for the evaluation
of improper integrals, the Fried \& Conte function, since it is basically
the complex complementary error function (also called the Faddeeva
function) and for $\mathscr{H}_{n}\left(z\right)$, via the modified
Bessel or Kummer hypergeometric functions.

\begin{subequations}
\label{eq:RKAP5:Z_scr-Rep_3}

However, for further mathematical properties it is convenient to derive
alternative and equivalent representations. Of particular importance
are representations in terms of double or triple integrals. In order
to obtain the latter from (\ref{eq:RKAP5:Z_scr-X_scr-Derivs}), we
revert the defining integrals of $Z\left(\zeta\right)$ and $\mathscr{H}_{n}\left(z\right)$
and affect the integration variables transformations $x\to\sqrt{\mu_{\perp}+\beta/\eta}x$
and $y\to\sqrt{\mu_{\parallel}+\beta/\eta}y$, thereby obtaining 
\begin{align}
\mathscr{Z}^{\left(\eta,\zeta,\boldsymbol{\mu}\right)}_{p_{1},p_{2},n}\left(\varrho,\varsigma\right) & =\frac{2}{\pi^{1/2}}\mathcal{N}^{\left(\eta,\zeta,\boldsymbol{\mu}\right)}\int^{\infty}_{0}dx\,\int^{\infty}_{-\infty}dy\,\frac{xJ^{2}_{n}\left(\nu x\right)}{y-\varsigma}e^{-\left(\mu_{\perp}x^{2}+\mu_{\parallel}y^{2}\right)}\nonumber \\
 & \times\int^{\infty}_{0}d\beta\,\frac{\beta^{\zeta-1}e^{-\left[1+\left(x^{2}+y^{2}\right)/\eta\right]\beta}}{\left(\mu_{\parallel}+\beta/\eta\right)^{p_{1}}\left(\mu_{\perp}+\beta/\eta\right)^{p_{2}-1}}\label{eq:RKAP5:ZsXsDR3-Z_scr}\\
\partial_{\varrho}\mathscr{Z}^{\left(\eta,\zeta,\boldsymbol{\mu}\right)}_{p_{1},p_{2},n}\left(\varrho,\varsigma\right) & =\frac{4\nu^{-1}}{\pi^{1/2}}\mathcal{N}^{\left(\eta,\zeta,\boldsymbol{\mu}\right)}\int^{\infty}_{0}dx\,\int^{\infty}_{-\infty}dy\,\frac{x^{2}J_{n}\left(\nu x\right)J^{\prime}_{n}\left(\nu x\right)}{y-\varsigma}e^{-\left(\mu_{\perp}x^{2}+\mu_{\parallel}y^{2}\right)}\nonumber \\
 & \times\int^{\infty}_{0}d\beta\,\frac{\beta^{\zeta-1}e^{-\left[1+\left(x^{2}+y^{2}\right)/\eta\right]\beta}}{\left(\mu_{\parallel}+\beta/\eta\right)^{p_{1}}\left(\mu_{\perp}+\beta/\eta\right)^{p_{2}-1}},\label{eq:RKAP5:ZsXsDR3-dZ_scrdr}
\end{align}
where $\nu=\sqrt{2\varrho}$\@. It is not necessary to repeat all
cases in (\ref{eq:RKAP5:Z_scr-X_scr-Derivs}), since the remaining
can be easily derived from the expressions above.
\end{subequations}

Finally, by inverting the procedure in the $y$-integral of (\ref{eq:RKAP5:ZsXsDR3-Z_scr}),
we obtain the third representation 
\begin{multline}
\mathscr{Z}^{\left(\eta,\zeta,\boldsymbol{\mu}\right)}_{p_{1},p_{2},n}\left(\varrho,\varsigma\right)=2\mathcal{N}^{\left(\eta,\zeta,\boldsymbol{\mu}\right)}\int^{\infty}_{0}dx\,xJ^{2}_{n}\left(\nu x\right)e^{-\mu_{\perp}x^{2}}\\
\times\int^{\infty}_{0}d\beta\,\frac{\beta^{\zeta-1}e^{-\left(1+x^{2}/\eta\right)\beta}Z\left(\tilde{\varsigma}\right)}{\left(\mu_{\parallel}+\beta/\eta\right)^{p_{1}}\left(\mu_{\perp}+\beta/\eta\right)^{p_{2}-1}}.\label{eq:RKAP5:Z_scr-Rep_4}
\end{multline}

Representations (\ref{eq:RKAP5:Z_scr-X_scr-Derivs}, \ref{eq:RKAP5:Z_scr-Rep_3}, and \ref{eq:RKAP5:Z_scr-Rep_4})
provide all necessary information to derive the ensuing mathematical
properties.

\subsubsection{Limit cases}

Important particular cases of the function $\mathscr{Z}^{\left(\eta,\zeta,\boldsymbol{\mu}\right)}_{p_{1},p_{2},n}\left(\varrho,\varsigma\right)$
can be derived from (\ref{eq:RKAP5:ZXD-Z_scr}, \ref{eq:RKAP5:ZsXsDR3-Z_scr}, and \ref{eq:RKAP5:Z_scr-Rep_4})\@.
The first one is the $\mu$-isotropic limit, 
\begin{equation}
\begin{aligned}\mathscr{Z}^{\left(\eta,\zeta,\mu\right)}_{p_{1},p_{2},n}\left(\varrho,\varsigma\right) & =\mathcal{N}^{\left(\eta,\zeta,\mu\right)}\int^{\infty}_{0}d\beta\,\frac{\beta^{\zeta-1}e^{-\beta}\mathscr{H}_{n}\left(\tilde{\varrho}\right)Z\left(\tilde{\varsigma}\right)}{\left(\mu+\beta/\eta\right)^{p_{1}+p_{2}}}\\
 & =\frac{2}{\pi^{1/2}}\mathcal{N}^{\left(\eta,\zeta,\mu\right)}\int^{\infty}_{0}dx\,\int^{\infty}_{-\infty}dy\,\frac{xJ^{2}_{n}\left(\nu x\right)}{y-\varsigma}e^{-\mu\left(x^{2}+y^{2}\right)}\\
 & \times\int^{\infty}_{0}d\beta\,\frac{\beta^{\zeta-1}e^{-\left[1+\left(x^{2}+y^{2}\right)/\eta\right]\beta}}{\left(\mu+\beta/\eta\right)^{p_{1}+p_{2}-1}}\\
 & =2\mathcal{N}^{\left(\eta,\zeta,\mu\right)}\int^{\infty}_{0}dx\,xJ^{2}_{n}\left(\nu x\right)e^{-\mu x^{2}}\int^{\infty}_{0}d\beta\,\frac{\beta^{\zeta-1}e^{-\left(1+x^{2}/\eta\right)\beta}Z\left(\tilde{\varsigma}\right)}{\left(\mu+\beta/\eta\right)^{p_{1}+p_{2}-1}},
\end{aligned}
\label{eq:Mu-isotropic_limit}
\end{equation}
where now $\mathcal{N}^{\left(\eta,\zeta,\mu\right)}$ is given by
(\ref{eq:Nc-mu_isotropic}) and 
\begin{align*}
\tilde{\varrho} & =\frac{\varrho}{\mu+\beta/\eta}, & \tilde{\varsigma} & =\sqrt{\mu+\frac{\beta}{\eta}}\varsigma.
\end{align*}
Obviously, $\mathscr{Z}^{\left(\eta,\zeta,\mu\right)}_{p_{1},p_{2},n}=\mathscr{Z}^{\left(\eta,\zeta,\mu\right)}_{p_{2},p_{1},n}$
or any other combination that results in the same value for $p_{1}+p_{2}$.

\begin{subequations}
\label{eq:SKD_limit}

The other important case is the SKD limit $\left(\mu=0\right)$, which
can be obtained from the second representation in (\ref{eq:Mu-isotropic_limit})
as
\begin{multline}
\mathscr{Z}^{\left(\eta,\zeta,0\right)}_{p_{1},p_{2},n}\left(\varrho,\varsigma\right)=\frac{2}{\pi^{1/2}}\frac{\Gamma\left(\zeta+1-p_{1}-p_{2}\right)}{\eta^{5/2-p_{1}-p_{2}}\Gamma\left(\zeta-\nicefrac{3}{2}\right)}\\
\times\int^{\infty}_{0}dx\,\int^{\infty}_{-\infty}dy\,\frac{xJ^{2}_{n}\left(\nu x\right)}{y-\varsigma}\left(1+\frac{x^{2}+y^{2}}{\eta}\right)^{-\left(\zeta+1-p_{1}-p_{2}\right)},\label{eq:SKDL-1}
\end{multline}
where we have used (\ref{eq:RKAP5:Superstatistics-ident} and \ref{eq:Nc-mu0}).

With the usual recipe for RKD $\longrightarrow$ SKD transition: $\eta=\kappa$,
$\zeta=\sigma=\kappa+\alpha$, and $p_{1}+p_{2}=2-\beta$, we then
have $\mathscr{Z}^{\left(\kappa,\sigma,0\right)}_{p_{1},p_{2},n}\left(\varrho,\varsigma\right)=\mathcal{Z}^{\left(\alpha,\beta\right)}_{n,\kappa}\left(\varrho,\varsigma\right)$,
where 
\begin{equation}
\mathcal{Z}^{\left(\alpha,\beta\right)}_{n,\kappa}\left(\varrho,\varsigma\right)=\frac{2}{\pi^{1/2}}\frac{\Gamma\left(\lambda-1\right)}{\kappa^{1/2+\beta}\Gamma\left(\sigma-\nicefrac{3}{2}\right)}\int^{\infty}_{0}dx\,\int^{\infty}_{-\infty}dy\,\frac{xJ^{2}_{n}\left(\nu x\right)}{y-\varsigma}\left(1+\frac{x^{2}+y^{2}}{\kappa}\right)^{-\left(\lambda-1\right)}\label{eq:SKDL-2}
\end{equation}
is the function defined by \citet{Gaelzer+16/06}, with $\lambda=\sigma+\beta$.
\end{subequations}

The final transition is the Maxwellian limit, obtained by taking $\kappa\to\infty$\@.
This limit can be obtained directly from (\ref{eq:SKDL-2}) and it
was shown by \citet{Gaelzer+16/06} that 
\begin{equation}
\lim_{\kappa\to\infty}\mathcal{Z}^{\left(\alpha,\beta\right)}_{n,\kappa}\left(\varrho,\varsigma\right)=\mathscr{H}_{n}\left(\varrho\right)Z\left(\varsigma\right).\label{eq:Maxwellian_limit}
\end{equation}

\subsection{Basic mathematical properties of $\mathscr{Z}^{\left(\eta,\zeta,\boldsymbol{\mu}\right)}_{p_{1},p_{2},n}\left(\varrho,\varsigma\right)$}\label{subsec:Z_scr-properties}

Here, some basic mathematical properties of the function $\mathscr{Z}^{\left(\eta,\zeta,\boldsymbol{\mu}\right)}_{p_{1},p_{2},n}\left(\varrho,\varsigma\right)$
will be derived, which can provide bases for analytical approximations
and  can also be implemented as computer code.

\subsubsection{Values at $\varsigma=0$ and $\varrho=0$}\label{subsec:RKAP5:Values-zero}

Let us start with $\varsigma=0$\@. Taking representations (\ref{eq:RKAP5:Z_scr-X_scr-Derivs}),
we know that $Z\left(0\right)=i\sqrt{\pi}$ and $Z^{\prime}\left(0\right)=-2$\@.
As a result, we obtain
\begin{align*}
\mathscr{Z}^{\left(\eta,\zeta,\boldsymbol{\mu}\right)}_{p_{1},p_{2},n}\left(\varrho,0\right) & =i\sqrt{\pi}\mathcal{M}^{\left(\eta,\zeta,\boldsymbol{\mu}\right)}_{p_{1},p_{2}}\mathscr{H}^{\left(\eta,\zeta,\boldsymbol{\mu}\right)}_{p_{1},p_{2},n}\left(\varrho\right)\\
\partial_{\varsigma}\mathscr{Z}^{\left(\eta,\zeta,\boldsymbol{\mu}\right)}_{p_{1},p_{2},n}\left(\varrho,0\right) & =-2\mathcal{M}^{\left(\eta,\zeta,\boldsymbol{\mu}\right)}_{p_{1}-\nicefrac{1}{2},p_{2}}\mathscr{H}^{\left(\eta,\zeta,\boldsymbol{\mu}\right)}_{p_{1}-\nicefrac{1}{2},p_{2},n}\left(\varrho\right)\\
\partial_{\varrho}\mathscr{Z}^{\left(\eta,\zeta,\boldsymbol{\mu}\right)}_{p_{1},p_{2},n}\left(\varrho,0\right) & =i\sqrt{\pi}\mathcal{M}^{\left(\eta,\zeta,\boldsymbol{\mu}\right)}_{p_{1},p_{2}}\mathscr{H}^{\left(\eta,\zeta,\boldsymbol{\mu}\right)\prime}_{p_{1},p_{2},n}\left(\varrho\right),
\end{align*}
where 
\begin{equation}
\mathcal{M}^{\left(\eta,\zeta,\boldsymbol{\mu}\right)}_{p_{1},p_{2}}=\frac{\Gamma\left(\zeta+1-p_{1}-p_{2}\right)}{\eta^{5/2-p_{1}-p_{2}}\Gamma\left(\zeta\right)\mathcal{U}^{\left(\eta,\zeta,\boldsymbol{\mu}\right)}}.\label{eq:RKAP5:M_cal}
\end{equation}

Function $\mathscr{H}^{\left(\eta,\zeta,\boldsymbol{\mu}\right)}_{p_{1},p_{2},n}\left(\varrho\right)$
is called here the \emph{anisotropic regularized kappa gyroradius
function}\@. Its definitions and properties are discussed below in
section \ref{subsec:ARKPGF}.

\begin{subequations}
\label{eq:RKAP5:ZX-rho0}

Now let us consider the case $\varrho=0$\@. From the expansion in
(\ref{eq:Gyroradius_function}), we obtain the limiting cases 
\begin{align*}
\mathscr{H}_{n}\left(0\right) & =\delta_{n,0}, & \mathscr{H}^{\prime}_{n}\left(0\right) & =-\delta_{n,0}+\frac{1}{2}\delta_{\left|n\right|,1}, & \left.\frac{n^{2}}{\varrho}\mathscr{H}_{n}\left(\tilde{\varrho}\right)\right|_{\varrho=0} & =\frac{1}{2}\frac{\delta_{\left|n\right|,1}}{\mu_{\perp}+\beta/\eta}.
\end{align*}
Employing these results in representations (\ref{eq:RKAP5:Z_scr-X_scr-Derivs}),
we obtain 
\begin{align}
\mathscr{Z}^{\left(\eta,\zeta,\boldsymbol{\mu}\right)}_{p_{1},p_{2},n}\left(0,\varsigma\right) & =\delta_{n,0}Z^{\left(\eta,\zeta,\boldsymbol{\mu}\right)}_{\kappa,p_{1}+1,p_{2}}\left(\varsigma\right)\\
\partial_{\varrho}\mathscr{Z}^{\left(\eta,\zeta,\boldsymbol{\mu}\right)}_{p_{1},p_{2},n}\left(0,\varsigma\right) & =\left(-\delta_{n,0}+\frac{1}{2}\delta_{\left|n\right|,1}\right)Z^{\left(\eta,\zeta,\boldsymbol{\mu}\right)}_{\kappa,p_{1}+1,p_{2}+1}\left(\varsigma\right)\\
\left.\frac{n^{2}}{\varrho}\mathscr{Z}^{\left(\eta,\zeta,\boldsymbol{\mu}\right)}_{p_{1},p_{2},n}\left(\varrho,\varsigma\right)\right|_{\varrho=0} & =\frac{1}{2}\delta_{\left|n\right|,1}Z^{\left(\eta,\zeta,\boldsymbol{\mu}\right)}_{\kappa,p_{1}+1,p_{2}+1}\left(\varsigma\right)\\
\mathscr{X}^{\left(\eta,\zeta,\boldsymbol{\mu}\right)}_{p_{1},p_{2},n}\left(0,\varsigma\right) & =\delta_{n,0}\mathcal{Z}^{\left(\eta,\zeta,\boldsymbol{\mu}\right)}_{\kappa,\boldsymbol{\theta},p_{1},p_{2}}\left(\varsigma\right)\\
\partial_{\varrho}\mathscr{X}^{\left(\eta,\zeta,\boldsymbol{\mu}\right)}_{p_{1},p_{2},n}\left(0,\varsigma\right) & =\left(-\delta_{n,0}+\frac{1}{2}\delta_{\left|n\right|,1}\right)\mathcal{Z}^{\left(\eta,\zeta,\boldsymbol{\mu}\right)}_{\kappa,\boldsymbol{\theta},p_{1},p_{2}+1}\left(\varsigma\right)\\
\left.\frac{n^{2}}{\varrho}\partial_{\varsigma}\mathscr{X}^{\left(\eta,\zeta,\boldsymbol{\mu}\right)}_{p_{1},p_{2},n}\left(\varrho,\varsigma\right)\right|_{\varrho=0} & =\frac{1}{2}\delta_{\left|n\right|,1}\mathcal{Z}^{\left(\eta,\zeta,\boldsymbol{\mu}\right)\prime}_{\kappa,\boldsymbol{\theta},p_{1},p_{2}+1}\left(\varsigma\right).
\end{align}
where now $Z^{\left(\eta,\zeta,\boldsymbol{\mu}\right)}_{\kappa,p_{1},p_{2}}\left(\varsigma\right)$
and $\mathcal{Z}^{\left(\eta,\zeta,\boldsymbol{\mu}\right)}_{\kappa,\boldsymbol{\theta},p_{1},p_{2}}\left(\varsigma\right)$
are the \emph{anisotropic regularized kappa plasma dispersion functions},
defined in section \ref{subsec:ARKPDF}\@. These results can also
be obtained from (\ref{eq:RKAP5:Gyroradius_expansion}).
\end{subequations}

\subsubsection{Recurrence relations}

Some recurrence relations can be derived from the representations,
such as a recurrence involving the derivative with respect to $\varsigma$\@.
Given (\ref{eq:RKAP5:ZXD-dZ_scr-dsigma}), and the equation 
\begin{equation}
Z^{\prime}\left(\zeta\right)=-2\left[1+\zeta Z\left(\zeta\right)\right],\label{eq:FC1:Diff-eq}
\end{equation}
and then identifying (\ref{eq:RKAP5:ZXD-Z_scr}, \ref{eq:RKAP5:M_cal}, \ref{eq:RKAP5:ARKGF-Rep_2}, and \ref{eq:Nc-gen}),
 we obtain
\begin{equation}
\partial_{\varsigma}\mathscr{Z}^{\left(\eta,\zeta,\boldsymbol{\mu}\right)}_{p_{1},p_{2},n}\left(\varrho,\varsigma\right)=-2\left[\mathcal{M}^{\left(\eta,\zeta,\boldsymbol{\mu}\right)}_{p_{1}-\nicefrac{1}{2},p_{2}}\mathscr{H}^{\left(\eta,\zeta,\boldsymbol{\mu}\right)}_{p_{1}-\nicefrac{1}{2},p_{2},n}\left(\varrho\right)+\varsigma\mathscr{Z}^{\left(\eta,\zeta,\boldsymbol{\mu}\right)}_{p_{1}-1,p_{2},n}\left(\varrho,\varsigma\right)\right].\label{eq:RKAP5:dZ_scr-dsig-RR}
\end{equation}

The $\mu=0$ limit becomes, from (\ref{eq:Mu-isotropic_limit} and \ref{eq:RKAP5:ARKGF-limits}),
\[
\partial_{\varsigma}\mathscr{Z}^{\left(\eta,\zeta,0\right)}_{p_{1},p_{2},n}\left(\varrho,\varsigma\right)=-2\left[\frac{\Gamma\left(\zeta+\nicefrac{3}{2}-p_{1}-p_{2}\right)}{\eta^{3-p_{1}-p_{2}}\Gamma\left(\zeta-\nicefrac{3}{2}\right)}\mathscr{H}^{\left(\eta,\zeta,0\right)}_{p_{1}-\nicefrac{1}{2},p_{2},n}\left(\varrho\right)+\varsigma\mathscr{Z}^{\left(\eta,\zeta,0\right)}_{p_{1}-1,p_{2},n}\left(\varrho,\varsigma\right)\right].
\]
Applying then the RKD $\to$ SKD recipe with (\ref{eq:SKD_limit} and \ref{eq:RKAP5:ARKGF-limits})
we obtain
\[
\partial_{\varsigma}\mathcal{Z}^{\left(\alpha,\beta\right)}_{n,\kappa}\left(\varrho,\varsigma\right)=-2\left[\frac{\Gamma\left(\lambda-\nicefrac{1}{2}\right)}{\kappa^{1+\beta}\Gamma\left(\sigma-\nicefrac{3}{2}\right)}\mathcal{H}^{\left(\alpha,\beta+\nicefrac{1}{2}\right)}_{n,\kappa}\left(\varrho\right)+\varsigma\mathcal{Z}^{\left(\alpha,\beta+1\right)}_{n,\kappa}\left(\varrho,\varsigma\right)\right],
\]
which is the recurrence relation obtained by \citet{GaelzerZiebell16/02}.

\subsubsection{Power series expansions}

Contrary to a standard kappa plasma, for the bRKd we can obtain power
series expansions for both arguments $\varrho$ and $\varsigma$\@.

\paragraph{\textbf{Expansion in powers of the gyroradius}}

\begin{subequations}
\label{eq:RKAP5:H_scr-series}

Representations (\ref{eq:RKAP5:Z_scr-X_scr-Derivs}) provide a practical
starting point for a series expansion in powers of the gyroradius
\[
k_{\perp}r_{L}=\frac{k_{\perp}\theta_{\perp}}{\sqrt{2}\Omega}=\sqrt{\varrho}.
\]
Taking the expansion for the function $\mathscr{H}_{n}\left(z\right)$
in (\ref{eq:Gyroradius_function}), we can also obtain compact expressions
for its derivative,\footnote{See, e.g., the appendices of \citet{GaelzerZiebell16/02,Gaelzer+16/06}.}
\begin{gather}
\mathscr{H}^{\prime}_{0}\left(z\right)=-\pFq 11\left({\nicefrac{3}{2}\atop 2};-2z\right)=-\sum^{\infty}_{k=0}\frac{\left(\nicefrac{3}{2}\right)_{k}}{\left(2\right)_{k}}\frac{\left(-2z\right)^{k}}{k!},\label{eq:HSS-H0p}\\
\begin{aligned}\left(\left|n\right|\geqslant1\right):\,\mathscr{H}^{\prime}_{n}\left(z\right) & =\frac{z^{\left|n\right|-1}}{2^{\left|n\right|}\left(\left|n\right|-1\right)!}\pFq 22\left({\left|n\right|+\nicefrac{1}{2},\left|n\right|+1\atop 2\left|n\right|+1,\left|n\right|};-2z\right)\\
 & =\frac{z^{\left|n\right|-1}}{2^{\left|n\right|}\left(\left|n\right|-1\right)!}\sum^{\infty}_{k=0}\frac{\left(\left|n\right|+\nicefrac{1}{2}\right)_{k}\left(\left|n\right|+1\right)_{k}}{\left(2\left|n\right|+1\right)_{k}\left(\left|n\right|\right)_{k}}\frac{\left(-2z\right)^{k}}{k!}.
\end{aligned}
\end{gather}
One can easily show that the result for $\mathscr{H}^{\prime}_{n}\left(z\right)$
can be transformed into (\ref{eq:HSS-H0p}).
\end{subequations}

\begin{subequations}
\label{eq:RKAP5:Gyroradius_expansion}

Inserting these expansions into (\ref{eq:RKAP5:Z_scr-X_scr-Derivs})
and then identifying the functions defined in section \ref{subsec:ARKPDF},
we obtain the gyroradius expansions
\begin{gather}
\begin{aligned}\mathscr{Z}^{\left(\eta,\zeta,\boldsymbol{\mu}\right)}_{p_{1},p_{2},n}\left(\varrho,\varsigma\right) & =\frac{\varrho^{\left|n\right|}}{2^{\left|n\right|}\left|n\right|!}\sum^{\infty}_{k=0}\frac{\left(\left|n\right|+\nicefrac{1}{2}\right)_{k}}{\left(2\left|n\right|+1\right)_{k}}\frac{\left(-2\varrho\right)^{k}}{k!}Z^{\left(\eta,\zeta,\boldsymbol{\mu}\right)}_{\kappa,p_{1}+1,p_{2}+\left|n\right|+k}\left(\varsigma\right)\\
 & =\frac{\varrho^{\left|n\right|}}{2^{\left|n\right|}\left|n\right|!}\left[\vphantom{\frac{\left|n\right|+\nicefrac{3}{2}}{\left|n\right|+1}}Z^{\left(\eta,\zeta,\boldsymbol{\mu}\right)}_{\kappa,p_{1}+1,p_{2}+\left|n\right|}\left(\varsigma\right)-\varrho Z^{\left(\eta,\zeta,\boldsymbol{\mu}\right)}_{\kappa,p_{1}+1,p_{2}+\left|n\right|+1}\left(\varsigma\right)\right.\\
 & \left.+\frac{1}{2}\frac{\left|n\right|+\nicefrac{3}{2}}{\left|n\right|+1}\varrho^{2}Z^{\left(\eta,\zeta,\boldsymbol{\mu}\right)}_{\kappa,p_{1}+1,p_{2}+\left|n\right|+k}\left(\varsigma\right)+\cdots\right]
\end{aligned}
\\
\begin{aligned}\partial_{\varrho}\mathscr{Z}^{\left(\eta,\zeta,\boldsymbol{\mu}\right)}_{p_{1},p_{2},0}\left(\varrho,\varsigma\right) & =-\sum^{\infty}_{k=0}\frac{\left(\nicefrac{3}{2}\right)_{k}}{\left(2\right)_{k}}\frac{\left(-2\varrho\right)^{k}}{k!}Z^{\left(\eta,\zeta,\boldsymbol{\mu}\right)}_{\kappa,p_{1}+1,p_{2}+1+k}\left(\varsigma\right)\\
 & =-Z^{\left(\eta,\zeta,\boldsymbol{\mu}\right)}_{\kappa,p_{1}+1,p_{2}+1}\left(\varsigma\right)+\frac{3}{2}\varrho Z^{\left(\eta,\zeta,\boldsymbol{\mu}\right)}_{\kappa,p_{1}+1,p_{2}+2}\left(\varsigma\right)-\frac{5}{4}\varrho^{2}Z^{\left(\eta,\zeta,\boldsymbol{\mu}\right)}_{\kappa,p_{1}+1,p_{2}+3}\left(\varsigma\right)+\cdots
\end{aligned}
\\
\begin{aligned}\partial_{\varrho}\mathscr{Z}^{\left(\eta,\zeta,\boldsymbol{\mu}\right)}_{p_{1},p_{2},n}\left(\varrho,\varsigma\right) & \stackrel{\left|n\right|\geqslant1}{=}\frac{\varrho^{\left|n\right|-1}}{2^{\left|n\right|}\left(\left|n\right|-1\right)!}\sum^{\infty}_{k=0}\frac{\left(\left|n\right|+\nicefrac{1}{2}\right)_{k}\left(\left|n\right|+1\right)_{k}}{\left(2\left|n\right|+1\right)_{k}\left(\left|n\right|\right)_{k}}\frac{\left(-2\varrho\right)^{k}}{k!}Z^{\left(\eta,\zeta,\boldsymbol{\mu}\right)}_{\kappa,p_{1}+1,p_{2}+\left|n\right|+k}\left(\varsigma\right)\\
 & =\frac{\varrho^{\left|n\right|-1}}{2^{\left|n\right|}\left(\left|n\right|-1\right)!}\left[Z^{\left(\eta,\zeta,\boldsymbol{\mu}\right)}_{\kappa,p_{1}+1,p_{2}+\left|n\right|}\left(\varsigma\right)-\frac{\left|n\right|+1}{\left|n\right|}\varrho Z^{\left(\eta,\zeta,\boldsymbol{\mu}\right)}_{\kappa,p_{1}+1,p_{2}+\left|n\right|+1}\left(\varsigma\right)\right.\\
 & \left.+\frac{1}{2}\frac{\left(\left|n\right|+\nicefrac{3}{2}\right)\left(\left|n\right|+2\right)}{\left|n\right|\left(\left|n\right|+1\right)}\varrho^{2}Z^{\left(\eta,\zeta,\boldsymbol{\mu}\right)}_{\kappa,p_{1}+1,p_{2}+\left|n\right|+2}\left(\varsigma\right)+\cdots\right]
\end{aligned}
\\
\begin{aligned}\mathscr{X}^{\left(\eta,\zeta,\boldsymbol{\mu}\right)}_{p_{1},p_{2},n}\left(\varrho,\varsigma\right) & =\frac{\varrho^{\left|n\right|}}{2^{\left|n\right|}\left|n\right|!}\sum^{\infty}_{k=0}\frac{\left(\left|n\right|+\nicefrac{1}{2}\right)_{k}}{\left(2\left|n\right|+1\right)_{k}}\frac{\left(-2\varrho\right)^{k}}{k!}\mathcal{Z}^{\left(\eta,\zeta,\boldsymbol{\mu}\right)}_{\kappa,\boldsymbol{\theta},p_{1},p_{2}+\left|n\right|+k}\left(\varsigma\right)\\
 & =\frac{\varrho^{\left|n\right|}}{2^{\left|n\right|}\left|n\right|!}\left[\vphantom{\frac{\left|n\right|+\nicefrac{3}{2}}{\left|n\right|+1}}\mathcal{Z}^{\left(\eta,\zeta,\boldsymbol{\mu}\right)}_{\kappa,\boldsymbol{\theta},p_{1},p_{2}+\left|n\right|}\left(\varsigma\right)-\varrho\mathcal{Z}^{\left(\eta,\zeta,\boldsymbol{\mu}\right)}_{\kappa,\boldsymbol{\theta},p_{1},p_{2}+\left|n\right|+1}\left(\varsigma\right)\right.\\
 & \left.+\frac{1}{2}\frac{\left|n\right|+\nicefrac{3}{2}}{\left|n\right|+1}\varrho^{2}\mathcal{Z}^{\left(\eta,\zeta,\boldsymbol{\mu}\right)}_{\kappa,\boldsymbol{\theta},p_{1},p_{2}+\left|n\right|+2}\left(\varsigma\right)+\cdots\right]
\end{aligned}
\\
\begin{aligned}\partial_{\varrho}\mathscr{X}^{\left(\eta,\zeta,\boldsymbol{\mu}\right)}_{p_{1},p_{2},0}\left(\varrho,\varsigma\right) & =-\sum^{\infty}_{k=0}\frac{\left(\nicefrac{3}{2}\right)_{k}}{\left(2\right)_{k}}\frac{\left(-2\varrho\right)^{k}}{k!}\mathcal{Z}^{\left(\eta,\zeta,\boldsymbol{\mu}\right)}_{\kappa,\boldsymbol{\theta},p_{1},p_{2}+1+k}\left(\varsigma\right)\\
 & =-\mathcal{Z}^{\left(\eta,\zeta,\boldsymbol{\mu}\right)}_{\kappa,\boldsymbol{\theta},p_{1},p_{2}+1}\left(\varsigma\right)+\frac{3}{2}\varrho\mathcal{Z}^{\left(\eta,\zeta,\boldsymbol{\mu}\right)}_{\kappa,\boldsymbol{\theta},p_{1},p_{2}+2}\left(\varsigma\right)-\frac{5}{4}\varrho^{2}\mathcal{Z}^{\left(\eta,\zeta,\boldsymbol{\mu}\right)}_{\kappa,\boldsymbol{\theta},p_{1},p_{2}+3}\left(\varsigma\right)+\cdots
\end{aligned}
\\
\begin{aligned}\partial_{\varrho}\mathscr{X}^{\left(\eta,\zeta,\boldsymbol{\mu}\right)}_{p_{1},p_{2},n}\left(\varrho,\varsigma\right) & \stackrel{\left|n\right|\geqslant1}{=}\frac{\varrho^{\left|n\right|-1}}{2^{\left|n\right|}\left(\left|n\right|-1\right)!}\sum^{\infty}_{k=0}\frac{\left(\left|n\right|+\nicefrac{1}{2}\right)_{k}\left(\left|n\right|+1\right)_{k}}{\left(2\left|n\right|+1\right)_{k}\left(\left|n\right|\right)_{k}}\frac{\left(-2\varrho\right)^{k}}{k!}\mathcal{Z}^{\left(\eta,\zeta,\boldsymbol{\mu}\right)}_{\kappa,\boldsymbol{\theta},p_{1},p_{2}+\left|n\right|+k}\left(\varsigma\right)\\
 & =\frac{\varrho^{\left|n\right|-1}}{2^{\left|n\right|}\left(\left|n\right|-1\right)!}\left[\mathcal{Z}^{\left(\eta,\zeta,\boldsymbol{\mu}\right)}_{\kappa,\boldsymbol{\theta},p_{1},p_{2}+\left|n\right|}\left(\varsigma\right)-\frac{\left|n\right|+1}{\left|n\right|}\varrho\mathcal{Z}^{\left(\eta,\zeta,\boldsymbol{\mu}\right)}_{\kappa,\boldsymbol{\theta},p_{1},p_{2}+\left|n\right|+1}\left(\varsigma\right)\right.\\
 & \left.+\frac{1}{2}\frac{\left(\left|n\right|+\nicefrac{3}{2}\right)\left(\left|n\right|+2\right)}{\left|n\right|\left(\left|n\right|+1\right)}\varrho^{2}\mathcal{Z}^{\left(\eta,\zeta,\boldsymbol{\mu}\right)}_{\kappa,\boldsymbol{\theta},p_{1},p_{2}+\left|n\right|+2}\left(\varsigma\right)+\cdots\right].
\end{aligned}
\end{gather}

The other cases in (\ref{eq:RKAP5:Z_scr-X_scr-Derivs}) are readily
derived from these. This kind of expansion is not readily obtainable
for a SKD plasma, as discussed at length by \citet{GaelzerZiebell14/12,GaelzerZiebell16/02,Gaelzer+16/06}.
\end{subequations}

\paragraph{\textbf{Expansion in powers of $\boldsymbol{\varsigma}$}}

Returning to (\ref{eq:RKAP5:Z_scr-X_scr-Derivs}), we now look for
a suitable power series expansion of the Fried \& Conte function.
The usual choice is the expansion of the complementary error function
\citep{Temme-Full-NIST10-1}, which then provides
\begin{align}
Z\left(\zeta\right) & =i\sqrt{\pi}e^{-\zeta^{2}}-2\zeta\sum^{\infty}_{k=0}\frac{\left(-\zeta^{2}\right)^{k}}{\left(\nicefrac{3}{2}\right)_{k}}, & Z^{\prime}\left(\zeta\right) & =-2i\sqrt{\pi}\zeta e^{-\zeta^{2}}-2\sum^{\infty}_{k=0}\frac{\left(-\zeta^{2}\right)^{k}}{\left(\nicefrac{1}{2}\right)_{k}}.\label{eq:RKAP5:Zfun-Series}
\end{align}

Then, identifying the functions $\mathscr{V}^{\left(\eta,\zeta,\boldsymbol{\mu}\right)}_{p_{1},p_{2},n}\left(\varrho,\varsigma\right)$
(defined in \ref{eq:RKAP5:V_scr}) and $\mathscr{H}^{\left(\eta,\zeta,\boldsymbol{\mu}\right)}_{p_{1},p_{2},n}\left(\varrho\right)$
(\ref{eq:ARKGF-Rep_1}), as well as (\ref{eq:RKAP5:M_cal}), we obtain
\begin{multline}
\mathscr{Z}^{\left(\eta,\zeta,\boldsymbol{\mu}\right)}_{p_{1},p_{2},n}\left(\varrho,\varsigma\right)=2i\sqrt{\pi}\mathscr{V}^{\left(\eta,\zeta,\boldsymbol{\mu}\right)}_{p_{1},p_{2},n}\left(\varrho,\varsigma\right)\\
-2\mathcal{M}^{\left(\eta,\zeta,\boldsymbol{\mu}\right)}_{p_{1}-\nicefrac{1}{2},p_{2}}\varsigma\sum^{\infty}_{k=0}\frac{\left(\zeta+\nicefrac{3}{2}-p_{1}-p_{2}\right)_{k}}{\left(\nicefrac{3}{2}\right)_{k}}\left(-\frac{\varsigma^{2}}{\eta}\right)^{k}\mathscr{H}^{\left(\eta,\zeta,\boldsymbol{\mu}\right)}_{p_{1}-\nicefrac{1}{2}-k,p_{2},n}\left(\varrho\right).\label{eq:RKAP5:Varsigma-series_expansion}
\end{multline}

It is interesting to derive the SKD limit of (\ref{eq:RKAP5:Varsigma-series_expansion}),
which is 
\begin{multline*}
\mathscr{Z}^{\left(\eta,\zeta,0\right)}_{p_{1},p_{2},n}\left(\varrho,\varsigma\right)=2i\sqrt{\pi}\mathscr{V}^{\left(\eta,\zeta,0\right)}_{p_{1},p_{2},n}\left(\varrho,\varsigma\right)-2\frac{\Gamma\left(\zeta+\nicefrac{3}{2}-p_{1}-p_{2}\right)}{\eta^{3/2-p_{1}-p_{2}}}\mathcal{N}^{\left(\eta,\zeta,0\right)}\varsigma\\
\times\sum^{\infty}_{k=0}\frac{\left(\zeta+\nicefrac{3}{2}-p_{1}-p_{2}\right)_{k}}{\left(\nicefrac{3}{2}\right)_{k}}\left(-\frac{\varsigma^{2}}{\eta}\right)^{k}\mathscr{H}^{\left(\eta,\zeta,0\right)}_{p_{1}-\nicefrac{1}{2}-k,p_{2},n}\left(\varrho\right).
\end{multline*}
At this point it is convenient to apply the RKD $\longrightarrow$
SKD transition recipe. From (\ref{eq:RKAP5:V_scr-Limits}, \ref{eq:RKAP5:ARKGFL-SKGF}, and \ref{eq:Nc-mu0}),
we obtain
\begin{multline*}
\mathscr{Z}^{\left(\kappa,\sigma,0\right)}_{p_{1},p_{2},n}\left(\varrho,\varsigma\right)=i\frac{\sqrt{\pi}\Gamma\left(\lambda-1\right)}{\kappa^{1/2+\beta}\Gamma\left(\sigma-\nicefrac{3}{2}\right)}\left(1+\frac{\varsigma^{2}}{\kappa}\right)^{-\left(\lambda-2\right)}\mathcal{H}^{\left(\alpha,\beta\right)}_{n,\kappa}\left[\varrho\left(1+\frac{\varsigma^{2}}{\kappa}\right)\right]\\
-\frac{2\Gamma\left(\lambda-\nicefrac{1}{2}\right)\varsigma}{\kappa^{\beta+1}\Gamma\left(\sigma-\nicefrac{3}{2}\right)}\sum^{\infty}_{k=0}\frac{\left(\lambda-\nicefrac{1}{2}\right)_{k}}{\left(\nicefrac{3}{2}\right)_{k}}\left(-\frac{\varsigma^{2}}{\kappa}\right)^{k}\mathcal{H}^{\left(\alpha,\beta-\nicefrac{1}{2}+k\right)}_{n,\kappa}\left(\varrho\right),
\end{multline*}
where, as usual, $\lambda=\sigma+\beta$\@. This is exactly the expansion
for the function $\mathcal{Z}^{\left(\alpha,\beta\right)}_{n,\kappa}\left(\varrho,\varsigma\right)$
derived by \citet{Gaelzer+16/06}.

\paragraph{\textbf{Asymptotic expansions}}

It really makes sense to solely pursue an expansion for $\left|\varsigma\right|\gg1$,
since a large Larmor radius result is usually not needed. The asymptotic
expansions of the Fried \& Conte function and its derivative are
\begin{align}
Z\left(\zeta\right) & \sim-\frac{1}{\zeta}\sum_{k=0}\frac{\left(\nicefrac{1}{2}\right)_{k}}{\zeta^{2k}}+i\epsilon\sqrt{\pi}e^{-\zeta^{2}}, & Z^{\prime}\left(\zeta\right) & \sim\sum_{k=0}\frac{\left(\nicefrac{3}{2}\right)_{k}}{\zeta^{2\left(k+1\right)}}-2i\epsilon\sqrt{\pi}\zeta e^{-\zeta^{2}},\label{eq:RKAP5:Z-Zp-asymptotic}
\end{align}
where $\epsilon=\left\{ 0\,\left(\Imag\, \zeta>0\right),\,1\,\left(\Imag\, \zeta=0\right),\,2\,\left(\Imag\, \zeta<0\right)\right\} $.

Inserting (\ref{eq:RKAP5:Z-Zp-asymptotic}) into (\ref{eq:RKAP5:Z_scr-Rep_4})
and identifying again (\ref{eq:RKAP5:M_cal}, \ref{eq:RKAP5:V_scr}, and \ref{eq:ARKGF-Rep_1}),
we obtain 
\begin{multline}
\mathscr{Z}^{\left(\eta,\zeta,\boldsymbol{\mu}\right)}_{p_{1},p_{2},n}\left(\varrho,\varsigma\right)=-\mathcal{M}^{\left(\eta,\zeta,\boldsymbol{\mu}\right)}_{p_{1}+\nicefrac{1}{2},p_{2}}\frac{1}{\varsigma}\sum_{k=0}\frac{\left(\nicefrac{1}{2}\right)_{k}\left(-\eta/\varsigma^{2}\right)^{k}}{\left(\nicefrac{1}{2}-\zeta+p_{1}+p_{2}\right)_{k}}\mathscr{H}^{\left(\eta,\zeta,\boldsymbol{\mu}\right)}_{p_{1}+\nicefrac{1}{2}+k,p_{2},n}\left(\varrho\right)\\
+2i\epsilon\sqrt{\pi}\mathscr{V}^{\left(\eta,\zeta,\boldsymbol{\mu}\right)}_{p_{1},p_{2},n}\left(\varrho,\varsigma\right).\label{eq:RKAP5:Varsigma-asymptotic_expansion}
\end{multline}

The straightforward SKD limit of (\ref{eq:RKAP5:Varsigma-asymptotic_expansion}),
with (\ref{eq:VSL-V_scr-H_cal} and \ref{eq:RKAP5:ARKGFL-SKGF}), is
\begin{multline*}
\mathscr{Z}^{\left(\kappa,\sigma,0\right)}_{p_{1},p_{2},n}\left(\varrho,\varsigma\right)=-\frac{\Gamma\left(\lambda-\nicefrac{3}{2}\right)}{\kappa^{\beta}\Gamma\left(\sigma-\nicefrac{3}{2}\right)}\frac{1}{\varsigma}\sum_{k=0}\frac{\left(\nicefrac{1}{2}\right)_{k}}{\left(\nicefrac{5}{2}-\lambda\right)_{k}}\left(-\frac{\kappa}{\varsigma^{2}}\right)^{k}\mathcal{H}^{\left(\alpha,\beta-\nicefrac{1}{2}-k\right)}_{n,\kappa}\left(\varrho\right)\\
+2i\epsilon\sqrt{\pi}\frac{\kappa^{-1/2-\beta}\Gamma\left(\lambda-1\right)}{2\Gamma\left(\sigma-\nicefrac{3}{2}\right)}\left(1+\frac{\varsigma^{2}}{\kappa}\right)^{-\left(\lambda-2\right)}\mathcal{H}^{\left(\alpha,\beta\right)}_{n,\kappa}\left[\varrho\left(1+\frac{\varsigma^{2}}{\kappa}\right)\right],
\end{multline*}
which is the asymptotic expansion of the function $\mathcal{Z}^{\left(\alpha,\beta\right)}_{n,\kappa}\left(\varrho,\varsigma\right)$
\citep{Gaelzer+16/06}.

\subsection{The function $\mathscr{V}^{\left(\eta,\zeta,\boldsymbol{\mu}\right)}_{p_{1},p_{2},n}\left(\varrho,\varsigma\right)$}

\begin{subequations}
\label{eq:RKAP5:V_scr-Expressions}

Another two-variables special function that appears in the analysis
of bRKd plasmas is 

\begin{multline}
\mathscr{V}^{\left(\eta,\zeta,\boldsymbol{\mu}\right)}_{p_{1},p_{2},n}\left(\varrho,\varsigma\right)=\mathcal{N}^{\left(\eta,\zeta,\boldsymbol{\mu}\right)}e^{-\mu_{\parallel}\varsigma^{2}}\int^{\infty}_{0}dx\,xJ^{2}_{n}\left(\nu x\right)e^{-\mu_{\perp}x^{2}}\\
\times\int^{\infty}_{0}d\beta\,\frac{\beta^{\zeta-1}e^{-\left[1+\left(x^{2}+\varsigma^{2}\right)/\eta\right]\beta}}{\left(\mu_{\parallel}+\beta/\eta\right)^{p_{1}}\left(\mu_{\perp}+\beta/\eta\right)^{p_{2}-1}}.\label{eq:RKAP5:V_scr}
\end{multline}

This double-integral definition can be reduced to single-integral
representations in two ways:
\begin{align}
\mathscr{V}^{\left(\eta,\zeta,\boldsymbol{\mu}\right)}_{p_{1},p_{2},n}\left(\varrho,\varsigma\right) & =\frac{1}{2}\mathcal{N}^{\left(\eta,\zeta,\boldsymbol{\mu}\right)}e^{-\mu_{\parallel}\varsigma^{2}}\int^{\infty}_{0}d\beta\,\frac{\beta^{\zeta-1}e^{-\left(1+\varsigma^{2}/\eta\right)\beta}\mathscr{H}_{n}\left(\tilde{\varrho}\right)}{\left(\mu_{\parallel}+\beta/\eta\right)^{p_{1}}\left(\mu_{\perp}+\beta/\eta\right)^{p_{2}}}\label{eq:V_scr-2}\\
\mathscr{V}^{\left(\eta,\zeta,\boldsymbol{\mu}\right)}_{p_{1},p_{2},n}\left(\varrho,\varsigma\right) & =\Gamma\left(\zeta\right)\eta^{p_{1}+p_{2}-1}\mathcal{N}^{\left(\eta,\zeta,\boldsymbol{\mu}\right)}e^{-\mu_{\parallel}\varsigma^{2}}\nonumber \\
 & \times\int^{\infty}_{0}dx\,xJ^{2}_{n}\left(\nu x\right)e^{-\mu_{\perp}x^{2}}\mathcal{U}^{\left(\eta,\zeta,\boldsymbol{\mu}\right)}_{p_{1}-\nicefrac{1}{2},p_{2}-2}\left(\sqrt{x^{2}+\varsigma^{2}}\right).\label{eq:V_scr-3}
\end{align}
\end{subequations}
Representation (\ref{eq:V_scr-2}) was is obtained using (\ref{eq:GMP01:Int-JJe})
in order to identify $\mathscr{H}_{n}\left(\tilde{\varrho}\right)$,
whereas representation (\ref{eq:V_scr-3}) was written in terms of
the function $\mathcal{U}^{\left(\eta,\zeta,\boldsymbol{\mu}\right)}_{\alpha_{1},\alpha_{2}}\left(z\right)$,
defined by (\ref{eq:RKAP5:U_cal_function-gen_def}).

\subsection*{Limit cases}

\begin{subequations}
\label{eq:RKAP5:V_scr-Limits}

From (\ref{eq:RKAP5:V_scr-Expressions}b,c), (\ref{eq:RKAP5:U_cal-fun-Mu_isotropic} and \ref{eq:Nc-mu_isotropic}),
we have the $\mu$-isotropic limits
\begin{align}
\mathscr{V}^{\left(\eta,\zeta,\mu\right)}_{p_{1},p_{2},n}\left(\varrho,\varsigma\right) & =\frac{1}{2}\mathcal{N}^{\left(\eta,\zeta,\mu\right)}e^{-\mu\varsigma^{2}}\int^{\infty}_{0}d\beta\,\frac{\beta^{\zeta-1}e^{-\left(1+\varsigma^{2}/\eta\right)\beta}}{\left(\mu+\beta/\eta\right)^{p_{1}+p_{2}}}\mathscr{H}_{n}\left(\frac{\varrho}{\mu+\beta/\eta}\right)\\
\mathscr{V}^{\left(\eta,\zeta,\mu\right)}_{p_{1},p_{2},n}\left(\varrho,\varsigma\right) & =\frac{\Gamma\left(\zeta\right)\left(\eta\mu\right)^{\zeta}}{\mu^{p_{1}+p_{2}-1}}\mathcal{N}^{\left(\eta,\zeta,\mu\right)}e^{-\mu\varsigma^{2}}\nonumber \\
 & \times\int^{\infty}_{0}dx\,xJ^{2}_{n}\left(\nu x\right)e^{-\mu x^{2}}U\left[\zeta,\zeta+2-p_{1}-p_{2},\eta\mu\left(1+\frac{x^{2}+\varsigma^{2}}{\eta}\right)\right].
\end{align}

The $\mu=0$ limit is obtained from (\ref{eq:V_scr-3}, \ref{eq:RKAP5:U_cal-fun-Mu0}, and \ref{eq:Nc-mu0})\@.
After introducing the variable $t=\left[1+\left(x^{2}+\varsigma^{2}\right)/\eta\right]\beta$
and using (\ref{eq:RKAP5:Superstatistics-ident}), we can apply the
RKD $\longrightarrow$ SKD transition recipe in order to obtain
\begin{equation}
\mathscr{V}^{\left(\kappa,\sigma,0\right)}_{p_{1},p_{2},n}\left(\varrho,\varsigma\right)=\frac{\kappa^{-1/2-\beta}\Gamma\left(\lambda-1\right)}{2\Gamma\left(\sigma-\nicefrac{3}{2}\right)}\left(1+\frac{\varsigma^{2}}{\kappa}\right)^{-\left(\lambda-2\right)}\mathcal{H}^{\left(\alpha,\beta\right)}_{n,\kappa}\left[\varrho\left(1+\frac{\varsigma^{2}}{\kappa}\right)\right],\label{eq:VSL-V_scr-H_cal}
\end{equation}
where $\mathcal{H}^{\left(\alpha,\beta\right)}_{n,\kappa}\left(z\right)$
is the SKD gyroradius function (\ref{eq:RKAP5:ARKGFL-SKGF}), with
$\lambda=\sigma+\beta$\@.
\end{subequations}

Further mathematical properties of $\mathscr{V}^{\left(\eta,\zeta,\boldsymbol{\mu}\right)}_{p_{1},p_{2},n}\left(\varrho,\varsigma\right)$
can be obtained from the results in section \ref{sec:Suprathermal_function_constant}
or from the same techniques employed in section \ref{subsec:Z_scr-properties}.

\subsection{The anisotropic regularized kappa plasma gyroradius function}\label{subsec:ARKPGF}

\begin{subequations}
\label{eq:RKAP5:Anisotropic_regularized_kappa_gyroradius_function}

The following function (and its derivative) appears in the formalism,
\begin{align}
\mathscr{H}^{\left(\eta,\zeta,\boldsymbol{\mu}\right)}_{p_{1},p_{2},n}\left(\varrho\right) & =\frac{2\eta^{1-p_{1}-p_{2}}}{\Gamma\left(\zeta+1-p_{1}-p_{2}\right)}\int^{\infty}_{0}dx\,xJ^{2}_{n}\left(\nu x\right)e^{-\mu_{\perp}x^{2}}\nonumber \\
 & \times\int^{\infty}_{0}\frac{d\beta\,\beta^{\zeta-1}e^{-\left(1+x^{2}/\eta\right)\beta}}{\left(\mu_{\parallel}+\beta/\eta\right)^{p_{1}}\left(\mu_{\perp}+\beta/\eta\right)^{p_{2}-1}}\label{eq:ARKGF-Rep_1}\\
\mathscr{H}^{\left(\eta,\zeta,\boldsymbol{\mu}\right)\prime}_{p_{1},p_{2},n}\left(\varrho\right) & =\frac{4\nu^{-1}\eta^{1-p_{1}-p_{2}}}{\Gamma\left(\zeta+1-p_{1}-p_{2}\right)}\int^{\infty}_{0}dx\,x^{2}J_{n}\left(\nu x\right)J^{\prime}_{n}\left(\nu x\right)e^{-\mu_{\perp}x^{2}}\nonumber \\
 & \times\int^{\infty}_{0}\frac{d\beta\,\beta^{\zeta-1}e^{-\left(1+x^{2}/\eta\right)\beta}}{\left(\mu_{\parallel}+\beta/\eta\right)^{p_{1}}\left(\mu_{\perp}+\beta/\eta\right)^{p_{2}-1}},
\end{align}
where $\nu=\sqrt{2\varrho}$.

The function also has the following alternative representations,
\begin{align}
\mathscr{H}^{\left(\eta,\zeta,\boldsymbol{\mu}\right)}_{p_{1},p_{2},n}\left(\varrho\right) & =\frac{\eta^{1-p_{1}-p_{2}}}{\Gamma\left(\zeta+1-p_{1}-p_{2}\right)}\int^{\infty}_{0}\frac{d\beta\,\beta^{\zeta-1}e^{-\beta}\mathscr{H}_{n}\left(\tilde{\varrho}\right)}{\left(\mu_{\parallel}+\beta/\eta\right)^{p_{1}}\left(\mu_{\perp}+\beta/\eta\right)^{p_{2}}},\label{eq:RKAP5:ARKGF-Rep_2}\\
\mathscr{H}^{\left(\eta,\zeta,\boldsymbol{\mu}\right)}_{p_{1},p_{2},n}\left(\varrho\right) & =\frac{2\Gamma\left(\zeta\right)}{\Gamma\left(\zeta+1-p_{1}-p_{2}\right)}\int^{\infty}_{0}dx\,xJ^{2}_{n}\left(\nu x\right)e^{-\mu_{\perp}x^{2}}\mathcal{U}^{\left(\eta,\zeta,\boldsymbol{\mu}\right)}_{p_{1}-\nicefrac{1}{2},p_{2}-2}\left(x\right).\label{eq:RKAP5:ARKGF-Rep_3}
\end{align}
The first one was obtained using (\ref{eq:GMP01:Int-JJe}), whereas
the last one was written in terms of the function $\mathcal{U}^{\left(\eta,\zeta,\boldsymbol{\mu}\right)}_{\alpha_{1},\alpha_{2}}\left(z\right)$,
defined by (\ref{eq:RKAP5:U_cal_function-gen_def}).
\end{subequations}

\subsubsection{Limit cases}

\begin{subequations}
\label{eq:RKAP5:ARKGF-limits}

The $\mu$-isotropic limit is obtained from (\ref{eq:RKAP5:ARKGF-Rep_3} and \ref{eq:RKAP5:U_cal-fun-Mu_isotropic}),
resulting
\begin{multline}
\mathscr{H}^{\left(\eta,\zeta,\mu\right)}_{p_{1},p_{2},n}\left(\varrho\right)=\frac{2\Gamma\left(\zeta\right)\left(\eta\mu\right)^{\zeta+1-p_{1}-p_{2}}}{\Gamma\left(\zeta+1-p_{1}-p_{2}\right)}\\
\times\int^{\infty}_{0}dx\,xJ^{2}_{n}\left(\nu x\right)e^{-\mu x^{2}}U\left(\zeta,\zeta+2-p_{1}-p_{2},\eta\mu\left(1+\frac{x^{2}}{\eta}\right)\right).\label{eq:RKAP5:ARKGF-mu_isotropic}
\end{multline}
For the $\mu=0$ limit, we return to (\ref{eq:RKAP5:ARKGF-Rep_3}),
evaluate using (\ref{eq:RKAP5:U_cal-fun-Mu0}) and obtain
\begin{equation}
\mathscr{H}^{\left(\eta,\zeta,0\right)}_{p_{1},p_{2},n}\left(\varrho\right)=2\int^{\infty}_{0}dx\,\frac{xJ^{2}_{n}\left(\nu x\right)}{\left(1+x^{2}/\eta\right)^{\zeta+1-p_{1}-p_{2}}}.
\end{equation}
With the usual recipe for RKD $\longrightarrow$ SKD transition, we
obtain 
\begin{equation}
\mathscr{H}^{\left(\kappa,\sigma,0\right)}_{p_{1},p_{2},n}\left(\varrho\right)=\mathcal{H}^{\left(\alpha,\beta\right)}_{n,\kappa}\left(\varrho\right)=2\int^{\infty}_{0}dx\,\frac{xJ^{2}_{n}\left(\nu x\right)}{\left(1+x^{2}/\kappa\right)^{\sigma+\beta-1}},\label{eq:RKAP5:ARKGFL-SKGF}
\end{equation}
where $\mathcal{H}^{\left(\alpha,\beta\right)}_{n,\kappa}\left(\varrho\right)$
is the SKD gyroradius function defined in \citet{GaelzerZiebell16/02}.
\end{subequations}

\subsubsection{Series expansion}

This function has a power series representation valid for any values
of $\eta$ and $\zeta$\@. Starting from representation (\ref{eq:RKAP5:ARKGF-Rep_2})
and using the series expansion (\ref{eq:Gyroradius_function}), one
readily obtains
\[
\mathscr{H}^{\left(\eta,\zeta,\boldsymbol{\mu}\right)}_{p_{1},p_{2},n}\left(\varrho\right)=\frac{\eta\Gamma\left(\zeta\right)}{\Gamma\left(\zeta+1-p_{1}-p_{2}\right)}\frac{\left(\eta\varrho\right)^{\left|n\right|}}{2^{\left|n\right|}\left|n\right|!}\sum^{\infty}_{k=0}\frac{\left(\nicefrac{1}{2}+\left|n\right|\right)_{k}}{\left(1+2\left|n\right|\right)_{k}}\mathcal{U}^{\left(\eta,\zeta,\boldsymbol{\mu}\right)}_{p_{1}-\nicefrac{1}{2},p_{2}+\left|n\right|+k-1}\frac{\left(-2\eta\varrho\right)^{k}}{k!},
\]
after identifying (\ref{eq:RKAP5:U_cal_const-gen-def})\@. If one
were to investigate the SKD limit of this expansion (applying \ref{eq:RKAP5:U_cal-fun-Mu0}), it would be equivalent to search for the power series expansion of
the function $\mathcal{H}^{\left(\alpha,\beta\right)}_{n,\kappa}\left(\varrho\right)$
in (\ref{eq:RKAP5:ARKGFL-SKGF})\@. As discussed by \citet{GaelzerZiebell16/02},
such power series only exists for noninteger $\kappa$ (or $\eta$). 

\subsection{The anisotropic regularized kappa plasma dispersion functions}\label{subsec:ARKPDF}

\begin{subequations}
\label{eq:RKAP5:Anisotropic_regularized_kappa_dispersion_functions}

The following functions also appear in the analysis, 
\begin{align}
Z^{\left(\eta,\zeta,\boldsymbol{\mu}\right)}_{\kappa,p_{1},p_{2}}\left(\varsigma\right) & =\mathcal{N}^{\left(\eta,\zeta,\boldsymbol{\mu}\right)}\int^{\infty}_{0}\frac{d\beta\,\beta^{\zeta-1}e^{-\beta}}{\left(\mu_{\parallel}+\beta/\eta\right)^{p_{1}-1}\left(\mu_{\perp}+\beta/\eta\right)^{p_{2}}}Z\left(\sqrt{\mu_{\parallel}+\frac{\beta}{\eta}}\varsigma\right)\label{eq:ARKDF-Z}\\
\mathcal{Z}^{\left(\eta,\zeta,\boldsymbol{\mu}\right)}_{\kappa,\boldsymbol{\theta},p_{1},p_{2}}\left(\varsigma\right) & =\mathcal{N}^{\left(\eta,\zeta,\boldsymbol{\mu}\right)}\int^{\infty}_{0}d\beta\,\frac{\mathscr{A}^{\mu}_{\theta}\left(\beta\right)\beta^{\zeta-1}e^{-\beta}}{\left(\mu_{\parallel}+\beta/\eta\right)^{p_{1}}\left(\mu_{\perp}+\beta/\eta\right)^{p_{2}}}Z\left(\sqrt{\mu_{\parallel}+\frac{\beta}{\eta}}\varsigma\right).
\end{align}

Another representation is obtained by introducing the definition of
the Fried \& Conte function (\ref{eq:Fried-Conte_function}) in (\ref{eq:ARKDF-Z}),
defining $y=\sqrt{\mu_{\parallel}+\beta/\eta}s$ and then identifying
(\ref{eq:RKAP5:U_cal_function-gen_def}), resulting
\begin{equation}
Z^{\left(\eta,\zeta,\boldsymbol{\mu}\right)}_{\kappa,p_{1},p_{2}}\left(\varsigma\right)=\frac{\Gamma\left(\zeta\right)}{\sqrt{\pi}}\eta^{p_{1}+p_{2}-1}\mathcal{N}^{\left(\eta,\zeta,\boldsymbol{\mu}\right)}\int^{\infty}_{-\infty}ds\,\frac{e^{-\mu_{\parallel}s^{2}}}{s-\varsigma}\mathcal{U}^{\left(\eta,\zeta,\boldsymbol{\mu}\right)}_{p_{1}-\nicefrac{3}{2},p_{2}-1}\left(s\right),\label{eq:RKAP5:ARKDFs-3}
\end{equation}
which is in the form of a Hilbert transform.

From the definition of $\mathscr{A}^{\mu}_{\theta}\left(\beta\right)$
we also have 
\begin{equation}
\begin{aligned}\mathcal{Z}^{\left(\eta,\zeta,\boldsymbol{\mu}\right)}_{\kappa,\boldsymbol{\theta},p_{1},p_{2}}\left(\varsigma\right) & =A_{\theta}Z^{\left(\eta,\zeta,\boldsymbol{\mu}\right)}_{\kappa,p_{1},p_{2}+1}\left(\varsigma\right)-\mu_{\parallel}A_{\mu}Z^{\left(\eta,\zeta,\boldsymbol{\mu}\right)}_{\kappa,p_{1}+1,p_{2}+1}\left(\varsigma\right)\\
 & =A_{\theta}Z^{\left(\eta,\zeta,\boldsymbol{\mu}\right)}_{\kappa,p_{1},p_{2}+1}\left(\varsigma\right)+\mu_{\perp}\mathcal{A}_{\mu}Z^{\left(\eta,\zeta,\boldsymbol{\mu}\right)}_{\kappa,p_{1}+1,p_{2}+1}\left(\varsigma\right).
\end{aligned}
\label{eq:RKAP5:ARKDF-Z_cal-Z}
\end{equation}
We also observe that 
\begin{equation}
Z^{\left(\eta,\zeta,\boldsymbol{\mu}\right)}_{\kappa,p_{1}+1,p_{2}}\left(\varsigma\right)-\mathcal{Z}^{\left(\eta,\zeta,\boldsymbol{\mu}\right)}_{\kappa,\boldsymbol{\theta},p_{1},p_{2}}\left(\varsigma\right)=\frac{\theta^{2}_{\perp}}{\theta^{2}_{\parallel}}Z^{\left(\eta,\zeta,\boldsymbol{\mu}\right)}_{\kappa,p_{1},p_{2}+1}\left(\varsigma\right).\label{eq:RKAP5:ARKDF-Z-Z_cal}
\end{equation}
\end{subequations}

These functions are generalizations of the functions $Z^{\left(\eta,\zeta,\boldsymbol{\mu}\right)}_{\kappa}\left(z\right)$
and $\mathcal{Z}^{\left(\eta,\zeta,\boldsymbol{\mu}\right)}_{\kappa,\boldsymbol{\theta}}\left(z\right)$
(and derivatives), 
\begin{align}
Z^{\left(\eta,\zeta,\boldsymbol{\mu}\right)}_{\kappa}\left(z\right) & =Z^{\left(\eta,\zeta,\boldsymbol{\mu}\right)}_{\kappa,1,1}\left(z\right), & \mathcal{Z}^{\left(\eta,\zeta,\boldsymbol{\mu}\right)}_{\kappa,\boldsymbol{\theta}}\left(z\right) & =\mathcal{Z}^{\left(\eta,\zeta,\boldsymbol{\mu}\right)}_{\kappa,\boldsymbol{\theta},1,1}\left(z\right),\label{eq:RKAP5:ZRKDFs-Parallel}
\end{align}
studied by \citet{Gaelzer+26/06-arxiv} for the analysis of parallel-propagating
waves in bRKd plasmas.

\subsubsection{Limit cases}

\begin{subequations}
\label{eq:RKAP5:ARKDFs-mu_isotropic}

From (\ref{eq:RKAP5:Anisotropic_regularized_kappa_dispersion_functions}),
the $\mu$-isotropic limits are
\begin{align}
Z^{\left(\eta,\zeta,\mu\right)}_{\kappa,p_{1},p_{2}}\left(\varsigma\right) & =\mathcal{N}^{\left(\eta,\zeta,\mu\right)}\int^{\infty}_{0}\frac{d\beta\,\beta^{\zeta-1}e^{-\beta}}{\left(\mu+\beta/\eta\right)^{p_{1}+p_{2}-1}}Z\left(\sqrt{\mu+\frac{\beta}{\eta}}\varsigma\right)\\
\mathcal{Z}^{\left(\eta,\zeta,\mu\right)}_{\kappa,\boldsymbol{\theta},p_{1},p_{2}}\left(\varsigma\right) & =A_{\theta}Z^{\left(\eta,\zeta,\mu\right)}_{\kappa,p_{1}+1,p_{2}}\left(\varsigma\right)=A_{\theta}Z^{\left(\eta,\zeta,\mu\right)}_{\kappa,p_{1},p_{2}+1}\left(\varsigma\right).
\end{align}

The $\mu=0$ limits are more easily identified using (\ref{eq:RKAP5:ARKDFs-3}, \ref{eq:RKAP5:U_cal-fun-Mu0}, and \ref{eq:Nc-mu0}),
resulting 
\[
Z^{\left(\eta,\zeta,0\right)}_{\kappa,p_{1},p_{2}}\left(\varsigma\right)=\frac{1}{\sqrt{\pi}}\frac{\Gamma\left(\zeta+1-p_{1}-p_{2}\right)}{\eta^{5/2-p_{1}-p_{2}}\Gamma\left(\zeta-\nicefrac{3}{2}\right)}\int^{\infty}_{-\infty}ds\,\frac{\left(1+s^{2}/\eta\right)^{-\left(\zeta+1-p_{1}-p_{2}\right)}}{s-\varsigma}.
\]
Finally, with the usual recipe for the RKD $\longrightarrow$ SKD
transition, we obtain 
\[
Z^{\left(\kappa,\sigma,0\right)}_{\kappa,p_{1},p_{2}}\left(\varsigma\right)=Z^{\left(\alpha,\beta\right)}_{\kappa}\left(\varsigma\right)=\frac{1}{\sqrt{\pi}}\frac{\Gamma\left(\lambda-1\right)}{\kappa^{1/2+\beta}\Gamma\left(\sigma-\nicefrac{3}{2}\right)}\int^{\infty}_{-\infty}ds\,\frac{\left(1+s^{2}/\kappa\right)^{-\left(\lambda-1\right)}}{s-\varsigma},
\]
where $\lambda=\sigma+\beta$ and where $Z^{\left(\alpha,\beta\right)}_{\kappa}\left(\varsigma\right)$
is the SKD dispersion function defined by \citet{GaelzerZiebell14/12,GaelzerZiebell16/02,Gaelzer+16/06}.
\end{subequations}

\subsubsection{Series expansion}

Returning to (\ref{eq:ARKDF-Z}), we insert expansions (\ref{eq:RKAP5:Zfun-Series})\@.
Then, with (\ref{eq:RKAP5:U_cal_function-gen_def} and \ref{eq:RKAP5:U_cal_const-gen-def})
we obtain 
\begin{multline}
Z^{\left(\eta,\zeta,\boldsymbol{\mu}\right)}_{\kappa,p_{1},p_{2}}\left(\varsigma\right)=i\sqrt{\pi}\frac{\Gamma\left(\zeta\right)\mathcal{N}^{\left(\eta,\zeta,\boldsymbol{\mu}\right)}}{\eta^{1-p_{1}-p_{2}}}e^{-\mu_{\parallel}\varsigma^{2}}\mathcal{U}^{\left(\eta,\zeta,\boldsymbol{\mu}\right)}_{p_{1}-\nicefrac{3}{2},p_{2}-1}\left(\varsigma\right)\\
-2\frac{\Gamma\left(\zeta\right)\mathcal{N}^{\left(\eta,\zeta,\boldsymbol{\mu}\right)}}{\eta^{3/2-p_{1}-p_{2}}}\varsigma\sum^{\infty}_{k=0}\frac{\left(-\varsigma^{2}/\eta\right)^{k}}{\left(\nicefrac{3}{2}\right)_{k}}\mathcal{U}^{\left(\eta,\zeta,\boldsymbol{\mu}\right)}_{p_{1}-2-k,p_{2}-1}.\label{eq:RKAP5:ARKDF-Series}
\end{multline}

From (\ref{eq:RKAP5:ARKDF-Series}, \ref{eq:RKAP5:U_cal-fun-Mu0}, and \ref{eq:Nc-mu0}),
the power series expansion after the RKD $\longrightarrow$ SKD transition
is
\begin{multline*}
Z^{\left(\kappa,\sigma,0\right)}_{\kappa,p_{1},p_{2}}\left(\varsigma\right)=i\frac{\sqrt{\pi}\Gamma\left(\lambda-1\right)}{\kappa^{1/2+\beta}\Gamma\left(\sigma-\nicefrac{3}{2}\right)}\left(1+\frac{\varsigma^{2}}{\kappa}\right)^{-\left(\lambda-1\right)}\\
-\frac{2\Gamma\left(\lambda-\nicefrac{1}{2}\right)}{\kappa^{1+\beta}\Gamma\left(\sigma-\nicefrac{3}{2}\right)}\varsigma\sum^{\infty}_{k=0}\frac{\left(\lambda-\nicefrac{1}{2}\right)_{k}}{\left(\nicefrac{3}{2}\right)_{k}}\left(-\frac{\varsigma^{2}}{\kappa}\right)^{k},
\end{multline*}
which is exactly the power series for the function $Z^{\left(\alpha,\beta\right)}_{\kappa}\left(\varsigma\right)$
obtained from Eq. (15) of \citet{GaelzerZiebell16/02}.

\subsubsection{Recurrence relations}

\begin{subequations}
\label{eq:Z-Z_cal-RR}

Some initial recurrence relations can be derived from the representations.
The first one will be a recurrence involving the derivative with respect
to $\varsigma$\@. Given (\ref{eq:FC1:Diff-eq}, \ref{eq:RKAP5:Anisotropic_regularized_kappa_dispersion_functions}, \ref{eq:RKAP5:U_cal_const-gen-def}, and \ref{eq:Nc-gen}),
we obtain
\begin{align}
Z^{\left(\eta,\zeta,\boldsymbol{\mu}\right)\prime}_{\kappa,p_{1},p_{2}}\left(\varsigma\right) & =-2\left[\eta^{p_{1}+p_{2}-3}\frac{\mathcal{U}^{\left(\eta,\zeta,\boldsymbol{\mu}\right)}_{p_{1}-2,p_{2}-1}}{\mathcal{U}^{\left(\eta,\zeta,\boldsymbol{\mu}\right)}}+\varsigma Z^{\left(\eta,\zeta,\boldsymbol{\mu}\right)}_{\kappa,p_{1}-1,p_{2}}\left(\varsigma\right)\right]\label{eq:RKAP5:dZ-dsig-RR}\\
\mathcal{Z}^{\left(\eta,\zeta,\boldsymbol{\mu}\right)\prime}_{\kappa,\boldsymbol{\theta},p_{1},p_{2}}\left(\varsigma\right) & =-2\left[\eta^{p_{1}+p_{2}-2}\mathcal{V}^{\left(\eta,\zeta,\boldsymbol{\mu}\right)}_{p_{1},p_{2}}+\varsigma\mathcal{Z}^{\left(\eta,\zeta,\boldsymbol{\mu}\right)}_{\kappa,\boldsymbol{\theta},p_{1}-1,p_{2}}\left(\varsigma\right)\right],\label{eq:RKAP5:dZ_cal-dsig-RR}
\end{align}
where 
\[
\mathcal{V}^{\left(\eta,\zeta,\boldsymbol{\mu}\right)}_{p_{1},p_{2}}=A_{\theta}\frac{\mathcal{U}^{\left(\eta,\zeta,\boldsymbol{\mu}\right)}_{p_{1}-2,p_{2}}}{\mathcal{U}^{\left(\eta,\zeta,\boldsymbol{\mu}\right)}}-\eta\mu_{\parallel}A_{\mu}\frac{\mathcal{U}^{\left(\eta,\zeta,\boldsymbol{\mu}\right)}_{p_{1}-1,p_{2}}}{\mathcal{U}^{\left(\eta,\zeta,\boldsymbol{\mu}\right)}}.
\]
\end{subequations}

\subsection{The suprathermal function $\mathcal{U}^{\left(\eta,\zeta,\boldsymbol{\mu}\right)}_{\alpha_{1},\alpha_{2}}\left(z\right)$
and the constant $\mathcal{U}^{\left(\eta,\zeta,\boldsymbol{\mu}\right)}_{\alpha_{1},\alpha_{2}}$}\label{sec:Suprathermal_function_constant}

The following function, 
\begin{equation}
\mathcal{U}^{\left(\eta,\zeta,\boldsymbol{\mu}\right)}_{\alpha_{1},\alpha_{2}}\left(z\right)=\frac{\eta^{-\left(\alpha_{1}+\alpha_{2}+\nicefrac{3}{2}\right)}}{\Gamma\left(\zeta\right)}\int^{\infty}_{0}\frac{d\beta\,\beta^{\zeta-1}e^{-\left(1+z^{2}/\eta\right)\beta}}{\left(\mu_{\parallel}+\beta/\eta\right)^{\alpha_{1}+\nicefrac{1}{2}}\left(\mu_{\perp}+\beta/\eta\right)^{\alpha_{2}+1}},\label{eq:RKAP5:U_cal_function-gen_def}
\end{equation}
frequently appears in the formalism of wave propagation in anisotropic
RKD plasmas. The function $\mathcal{U}^{\left(\eta,\zeta,\boldsymbol{\mu}\right)}_{\alpha_{1},\alpha_{2}}\left(z\right)$
is defined for complex $z$ and has several representations in terms
of the Tricomi confluent hypergeometric and gamma functions, as shown
below.

Associated to the function, there is also the constant 
\begin{equation}
\mathcal{U}^{\left(\eta,\zeta,\boldsymbol{\mu}\right)}_{\alpha_{1},\alpha_{2}}=\mathcal{U}^{\left(\eta,\zeta,\boldsymbol{\mu}\right)}_{\alpha_{1},\alpha_{2}}\left(0\right).\label{eq:RKAP5:U_cal_const-gen-def}
\end{equation}

Some representations and particular cases of the function and the
associated constant will be presented in the following sections.

\subsubsection{Euler-type integral representation}

Let us start from the identity 
\[
\frac{1}{A^{\gamma_{1}}_{1}A^{\gamma_{2}}_{2}}=\frac{\Gamma\left(\gamma_{1}+\gamma_{2}\right)}{\Gamma\left(\gamma_{1}\right)\Gamma\left(\gamma_{2}\right)}\int^{1}_{0}dx_{1}\int^{1}_{0}dx_{2}\frac{x^{\gamma_{1}-1}_{1}x^{\gamma_{2}-1}_{2}\delta\left(1-x_{1}-x_{2}\right)}{\left(x_{1}A_{1}+x_{2}A_{2}\right)^{\gamma_{1}+\gamma_{2}}}\quad\left(\gamma_{1},\gamma_{2}>0\right),
\]
known as Feynman parametrization, and write the function (\ref{eq:RKAP5:U_cal_function-gen_def})
as 
\begin{equation}
\mathcal{U}^{\left(\eta,\zeta,\boldsymbol{\mu}\right)}_{\alpha_{1},\alpha_{2}}\left(z\right)=\frac{\eta^{-\left(\alpha_{1}+\alpha_{2}+\nicefrac{3}{2}\right)}}{\Gamma\left(\zeta\right)}\int^{\infty}_{0}\frac{d\beta\,\beta^{\zeta-1}e^{-\left(1+z^{2}/\eta\right)\beta}}{\left(\mu_{+}+\beta/\eta\right)^{\alpha_{+}}\left(\mu_{-}+\beta/\eta\right)^{\alpha_{-}}},\label{eq:RKAP5:U_cal_function-gen-2}
\end{equation}
where 
\begin{align*}
\mu_{+} & =\max\left(\mu_{\parallel},\mu_{\perp}\right), & \mu_{-} & =\min\left(\mu_{\parallel},\mu_{\perp}\right), & \alpha_{+(-)} & =\begin{cases}
\alpha_{1}+\nicefrac{1}{2}, & \mu_{+(-)}=\mu_{\parallel}\\
\alpha_{2}+1, & \mu_{+(-)}=\mu_{\perp}.
\end{cases}
\end{align*}
 Then, 
\begin{align*}
\mathcal{U}^{\left(\eta,\zeta,\boldsymbol{\mu}\right)}_{\alpha_{1},\alpha_{2}}\left(z\right) & =\frac{\Gamma\left(\alpha_{1}+\alpha_{2}+\nicefrac{3}{2}\right)\left(\mu_{+}\eta\right)^{-\left(\alpha_{1}+\alpha_{2}+\nicefrac{3}{2}\right)}}{\Gamma\left(\zeta\right)\Gamma\left(\alpha_{1}+\nicefrac{1}{2}\right)\Gamma\left(\alpha_{2}+1\right)}\\
 & \times\int^{1}_{0}dx_{1}\int^{1}_{0}dx_{2}\,x^{\alpha_{+}-1}_{1}x^{\alpha_{-}-1}_{2}\delta\left(1-x_{1}-x_{2}\right)\\
 & \times\int^{\infty}_{0}\frac{d\beta\,\beta^{\zeta-1}e^{-\left(1+z^{2}/\eta\right)\beta}}{\left[x_{1}+\left(1-A_{\pm}\right)x_{2}+\left(x_{1}+x_{2}\right)\beta/\mu_{+}\eta\right]^{\alpha_{1}+\alpha_{2}+\nicefrac{3}{2}}},
\end{align*}
where 
\[
A_{\pm}=1-\frac{\mu_{-}}{\mu_{+}}\quad\left(0\leqslant A_{\pm}\leqslant1\right).
\]

Then, defining 
\[
\beta=\mu_{+}\eta\left(1-A_{\pm}x\right)t,
\]
and using the following representation of the Tricomi confluent hypergeometric
function $U\left(a,b,z\right)$ \citep{Daalhuis-Full-NIST10a}, \begin{subequations}\label{eq:Tricomi_identities}
\begin{equation}
U\left(a,b,z\right)=\frac{1}{\Gamma\left(a\right)}\int^{\infty}_{0}dt\,t^{a-1}\left(1+t\right)^{b-a-1}e^{-zt}\:\left(\Real\,  a>0,\left|\arg\left(z\right)\right|<\frac{\pi}{2}\right),\label{eq:RKAP5:Tricomi-Integ_rep-1}
\end{equation}
and the associated Kummer transformation
\begin{equation}
U\left(a,b,z\right)=z^{1-b}U\left(a-b+1,2-b,z\right),\label{eq:RKAP2:Kummer-transf}
\end{equation}
\end{subequations}we obtain the following equivalent representations
as Euler-type integrals, \begin{subequations}\label{eq:RKAP5:U_cal_function-Euler_integrals}
\begin{align}
\mathcal{U}^{\left(\eta,\zeta,\boldsymbol{\mu}\right)}_{\alpha_{1},\alpha_{2}}\left(z\right) & =\frac{\Gamma\left(\alpha_{1}+\alpha_{2}+\nicefrac{3}{2}\right)}{\Gamma\left(\alpha_{1}+\nicefrac{1}{2}\right)\Gamma\left(\alpha_{2}+1\right)}\left(\eta\mu_{+}\right)^{\zeta-\alpha_{1}-\alpha_{2}-\nicefrac{3}{2}}\int^{1}_{0}dx\frac{x^{\alpha_{-}-1}\left(1-x\right)^{\alpha_{+}-1}}{\left(1-A_{\pm}x\right)^{\alpha_{1}+\alpha_{2}-\zeta+3/2}}\nonumber \\
 & \times U\left[\zeta,\zeta-\alpha_{1}-\alpha_{2}-\frac{1}{2},\eta\mu_{+}\left(1-A_{\pm}x\right)\left(1+\frac{z^{2}}{\eta}\right)\right],\label{eq:RKAP5:UcFEI-1}\\
\mathcal{U}^{\left(\eta,\zeta,\boldsymbol{\mu}\right)}_{\alpha_{1},\alpha_{2}}\left(z\right) & =\frac{\Gamma\left(\alpha_{1}+\alpha_{2}+\nicefrac{3}{2}\right)}{\Gamma\left(\alpha_{1}+\nicefrac{1}{2}\right)\Gamma\left(\alpha_{2}+1\right)}\left(1+\frac{z^{2}}{\eta}\right)^{\alpha_{1}+\alpha_{2}-\zeta+\nicefrac{3}{2}}\int^{1}_{0}dx\,x^{\alpha_{-}-1}\left(1-x\right)^{\alpha_{+}-1}\nonumber \\
 & \times U\left[\alpha_{1}+\alpha_{2}+\frac{3}{2},\alpha_{1}+\alpha_{2}-\zeta+\frac{5}{2},\eta\mu_{+}\left(1-A_{\pm}x\right)\left(1+\frac{z^{2}}{\eta}\right)\right]\label{eq:RKAP5:UcFEI-2}
\end{align}
which are valid for $\alpha_{1}>-\nicefrac{1}{2}$ and $\alpha_{2}>-1$\@.
In this representation, when $\zeta$ is noninteger, the function
has two branch lines, running along the imaginary axis with $\Imag\,  z\geqslant\sqrt{\eta}$
and $\Imag\,  z\leqslant-\sqrt{\eta}$.\end{subequations}

The Euler-type integral representations of the constant $\mathcal{U}^{\left(\eta,\zeta,\boldsymbol{\mu}\right)}_{\alpha_{1},\alpha_{2}}$
are obtained from (\ref{eq:RKAP5:U_cal_function-Euler_integrals})
at $z=0$.

\subsubsection{Power series representation}

Let us now return to (\ref{eq:RKAP5:U_cal_function-gen-2}) and first
define the new variable $\beta=\eta\mu_{+}t$, obtaining
\[
\mathcal{U}^{\left(\eta,\zeta,\boldsymbol{\mu}\right)}_{\alpha_{1},\alpha_{2}}\left(z\right)=\frac{1}{\Gamma\left(\zeta\right)}\left(\eta\mu_{+}\right)^{\zeta-\alpha_{1}-\alpha_{2}-\nicefrac{3}{2}}\int^{\infty}_{0}\frac{dt\,t^{\zeta-1}e^{-\left(1+z^{2}/\eta\right)\eta\mu_{+}t}}{\left(1+t\right)^{\alpha_{+}}\left(1+t-A_{\pm}\right)^{\alpha_{-}}}.
\]
Given now the binomial theorem \citep{GradshteynRyzhik07}
\[
\frac{1}{\left(1+t-A_{\lessgtr}\right)^{\alpha_{-}}}=\frac{1}{\left(1+t\right)^{\alpha_{-}}}\sum^{\infty}_{\ell=0}\frac{\left(\alpha_{-}\right)_{\ell}}{\ell!}\left(\frac{A_{\pm}}{1+t}\right)^{\ell},
\]
we obtain 
\begin{equation}
\mathcal{U}^{\left(\eta,\zeta,\boldsymbol{\mu}\right)}_{\alpha_{1},\alpha_{2}}\left(z\right)=\left(\eta\mu_{+}\right)^{\zeta-\alpha_{1}-\alpha_{2}-\nicefrac{3}{2}}\sum^{\infty}_{\ell=0}\frac{\left(\alpha_{-}\right)_{\ell}}{\ell!}U\left[\zeta,\zeta-\alpha_{1}-\alpha_{2}-\frac{1}{2}-\ell,\eta\mu_{+}\left(1+\frac{z^{2}}{\eta}\right)\right]A^{\ell}_{\pm},\label{eq:RKAP5:U_cal_function-Series}
\end{equation}
where we have used (\ref{eq:RKAP5:Tricomi-Integ_rep-1}) again. This
expansion is valid for any $\left\{ \alpha_{1},\alpha_{2}\right\} $.

The power series expansion of the constant $\mathcal{U}^{\left(\eta,\zeta,\boldsymbol{\mu}\right)}_{\alpha_{1},\alpha_{2}}$
is obtained from (\ref{eq:RKAP5:U_cal_function-Series}) at $z=0$.

\subsubsection{Limit cases}

Important particular cases of $\mathcal{U}^{\left(\eta,\zeta,\boldsymbol{\mu}\right)}_{\alpha_{1},\alpha_{2}}\left(z\right)$
and $\mathcal{U}^{\left(\eta,\zeta,\boldsymbol{\mu}\right)}_{\alpha_{1},\alpha_{2}}=\mathcal{U}^{\left(\eta,\zeta,\boldsymbol{\mu}\right)}_{\alpha_{1},\alpha_{2}}\left(0\right)$
are the following.

\paragraph{\textbf{Isotropic $\boldsymbol{\mu}$ case}}

Returning to (\ref{eq:RKAP5:U_cal_function-gen_def}) with $\mu_{\perp}=\mu_{\parallel}=\mu$,
and then defining $\beta=\eta\mu t$ and using (\ref{eq:Tricomi_identities}),
we obtain
\begin{equation}
\begin{aligned}\mathcal{U}^{\left(\eta,\zeta,\mu\right)}_{\alpha_{1},\alpha_{2}}\left(z\right) & =\left(\eta\mu\right)^{\zeta-\alpha_{1}-\alpha_{2}-\nicefrac{3}{2}}U\left[\zeta,\zeta-\alpha_{1}-\alpha_{2}-\frac{1}{2},\eta\mu\left(1+\frac{z^{2}}{\eta}\right)\right]\\
 & =\left(1+\frac{z^{2}}{\eta}\right)^{\alpha_{1}+\alpha_{2}-\zeta+\nicefrac{3}{2}}U\left[\alpha_{1}+\alpha_{2}+\frac{3}{2},\alpha_{1}+\alpha_{2}-\zeta+\frac{5}{2},\eta\mu\left(1+\frac{z^{2}}{\eta}\right)\right].
\end{aligned}
\label{eq:RKAP5:U_cal-fun-Mu_isotropic}
\end{equation}
 The same result could be obtained from either (\ref{eq:RKAP5:U_cal_function-Euler_integrals})
or (\ref{eq:RKAP5:U_cal_function-Series}).

\paragraph{\textbf{SKD limit}}

Returning again to (\ref{eq:RKAP5:U_cal_function-gen_def}) and taking
the case $\mu_{\perp}=\mu_{\parallel}=0$, 
\[
\mathcal{U}^{\left(\eta,\zeta,0\right)}_{\alpha_{1},\alpha_{2}}\left(z\right)=\frac{1}{\Gamma\left(\zeta\right)}\int^{\infty}_{0}d\beta\,\beta^{\zeta-\alpha_{1}-\alpha_{2}-\nicefrac{5}{2}}e^{-\left(1+z^{2}/\eta\right)\beta}.
\]
Using now identity (\ref{eq:RKAP5:Superstatistics-ident}), we obtain
\begin{equation}
\mathcal{U}^{\left(\eta,\zeta,0\right)}_{\alpha_{1},\alpha_{2}}\left(z\right)=\frac{\Gamma\left(\zeta-\alpha_{1}-\alpha_{2}-\nicefrac{3}{2}\right)}{\Gamma\left(\zeta\right)}\left(1+\frac{z^{2}}{\eta}\right)^{-\left(\zeta-\alpha_{1}-\alpha_{2}-\nicefrac{3}{2}\right)}.\label{eq:RKAP5:U_cal-fun-Mu0}
\end{equation}

\subsubsection{Normalization constants}

\begin{subequations}
\label{eq:Norm_constants}

The normalization constant in (\ref{eq:RKAP3:RKD-anisotropic-beam-generalized})
is written in terms of $\mathcal{U}^{\left(\eta,\zeta,\boldsymbol{\mu}\right)}_{\alpha_{1},\alpha_{2}}$
as 
\begin{equation}
N^{(\eta,\zeta,\boldsymbol{\mu})}_{\kappa,\boldsymbol{\theta}}=\left[\left(\pi\eta\right)^{3/2}\theta_{\parallel}\theta^{2}_{\perp}\mathcal{U}^{\left(\eta,\zeta,\boldsymbol{\mu}\right)}\right]^{-1},\label{eq:Norm_const}
\end{equation}
where $\mathcal{U}^{\left(\eta,\zeta,\boldsymbol{\mu}\right)}=\mathcal{U}^{\left(\eta,\zeta,\boldsymbol{\mu}\right)}_{0,0}$\@.
The SKD limit of $N^{(\eta,\zeta,\boldsymbol{\mu})}_{\kappa,\boldsymbol{\theta}}$
is of particular importance and is obtained from (\ref{eq:RKAP5:U_cal-fun-Mu0}),
resulting
\begin{equation}
N^{(\eta,\zeta,0)}_{\kappa,\boldsymbol{\theta}}=\left[\left(\pi\eta\right)^{3/2}\theta_{\parallel}\theta^{2}_{\perp}\frac{\Gamma\left(\zeta-\nicefrac{3}{2}\right)}{\Gamma\left(\zeta\right)}\right]^{-1}\quad\left(\zeta>\frac{3}{2}\right).\label{eq:Norm_const-SKD}
\end{equation}
\end{subequations}

\begin{subequations}
\label{eq:RKAP5:N_cal}

In (\ref{eq:RKAP5:Z_scr-X_scr-Derivs}) we have also introduced the
normalization factor 
\begin{equation}
\mathcal{N}^{\left(\eta,\zeta,\boldsymbol{\mu}\right)}=\left[\eta^{3/2}\Gamma\left(\zeta\right)\mathcal{U}^{\left(\eta,\zeta,\boldsymbol{\mu}\right)}\right]^{-1}.\label{eq:Nc-gen}
\end{equation}
The $\mu$-isotropic limit is given by (\ref{eq:RKAP5:U_cal-fun-Mu_isotropic}),
resulting
\begin{equation}
\mathcal{N}^{\left(\eta,\zeta,\mu\right)}=\left[\Gamma\left(\zeta\right)\eta^{3/2}\left(\eta\mu\right)^{\zeta-3/2}U\left(\zeta,\zeta-\frac{1}{2},\eta\mu\right)\right]^{-1}=\left[\Gamma\left(\zeta\right)\eta^{3/2}U\left(\frac{3}{2},\frac{5}{2}-\zeta,\eta\mu\right)\right]^{-1},\label{eq:Nc-mu_isotropic}
\end{equation}
whereas the $\mu=0$ limit is given by (\ref{eq:RKAP5:U_cal-fun-Mu0}),
resulting
\begin{equation}
\mathcal{N}^{\left(\eta,\zeta,0\right)}=\left[\eta^{3/2}\Gamma\left(\zeta-\nicefrac{3}{2}\right)\right]^{-1}\quad\left(\zeta>\frac{3}{2}\right).\label{eq:Nc-mu0}
\end{equation}
\end{subequations}

\section{Summary}

In this work, we provide for the first time a theoretical framework
for the analysis of the propagation of waves and linear wave-particle
interaction in plasmas described by fully anisotropic regularized
Kappa with drift distributions. The formalism is valid for  any propagation
angle, wave frequency range and wave polarization, since the complete
susceptibility tensor for a bRKd plasma is derived, and generalizes
a previous contribution that was restricted to propagation along the
ambient magnetic field.

In the course of the derivation of the tensor components, new plasma
dispersion and gyroradius functions were defined, which generalize
the corresponding functions previously defined for the parallel case
and which also have as limiting cases the known functions for a standard
Kappa and Maxwellian distributions. Beside the basic definitions,
alternative and equivalent representations were also obtained, as
well as several mathematical properties that enable the immediate
implementation of the theoretical framework in dispersion equation
solvers or the derivation of approximate expressions for the dispersion
relations and growth/damping rate coefficients.

Among the plenty possible applications of our formulation, we are
planning the implementation of the present results in a new comprehensive
dispersion equation solver, that will support a wide range of different
plasma populations, modelled by various models of velocity distribution
functions, among which the bRKd will be included. In this way, the
space plasma physics community will have access to a new tool that
complements other fully numerical approaches.

\section*{Acknowledgements}

We gratefully acknowledge financial support from the Deutsche Forschungsgesellschaft (DFG) for a Mercator Fellowship for R.G\@.  The research was funded within the WEAVE scheme by the DFG project No. FI 706/31-1, and the Belgian FWO-Vlaanderen, project G002523N.

\appendix

\section{Sums of Bessel functions}\label{sec:Sums-of-Bessel}

From Neumann's addition theorem \citep{OlverMaximon-Full-NIST10}
\[
J_{\nu}\left(u\pm v\right)=\sum^{\infty}_{k\to-\infty}J_{\nu\mp k}\left(u\right)J_{k}\left(v\right),
\]
and the property $J_{-n}\left(z\right)=\left(-\right)^{n}J_{n}\left(z\right)$,
one presently derives the following identities
\begin{align*}
\sum^{\infty}_{n\to-\infty}J^{2}_{n}\left(z\right) & =1, & \sum^{\infty}_{n\to-\infty}nJ^{2}_{n}\left(z\right) & =0, & \sum^{\infty}_{n\to-\infty}J_{n}\left(z\right)J^{\prime}_{n}\left(z\right) & =0.
\end{align*}

\section{Proof of eq. (\ref{eq:KAP7:KPSF2V-Y-dZ})}\label{sec:Proof-of-eq_9}

We will start by considering the integral 
\[
I_{x}=\int^{\infty}_{0}dx\,x^{3}J^{\prime2}_{n}\left(\alpha x\right)e^{-\rho^{2}x^{2}}.
\]
By first integrating by parts and then using Bessel's equation \citep{OlverMaximon-Full-NIST10},
we obtain 
\[
I_{x}=\frac{n^{2}}{\alpha\rho^{2}}\int^{\infty}_{0}dx\,J_{n}\left(\alpha x\right)J^{\prime}_{n}\left(\alpha x\right)e^{-\rho^{2}x^{2}}-\frac{\alpha}{\rho^{2}}\int^{\infty}_{0}dx\,x^{2}J_{n}\left(\alpha x\right)J^{\prime}_{n}\left(\alpha x\right)e^{-\rho^{2}x^{2}}.
\]
However, by integrating by parts the first integral, we obtain a first
identity, 
\begin{multline}
\int^{\infty}_{0}dx\,x^{3}J^{\prime2}_{n}\left(\alpha x\right)e^{-\rho^{2}x^{2}}=\frac{n^{2}}{\alpha^{2}}\int^{\infty}_{0}dx\,xJ^{2}_{n}\left(\alpha x\right)e^{-\rho^{2}x^{2}}\\
-\frac{\alpha}{\rho^{2}}\int^{\infty}_{0}dx\,x^{2}J_{n}\left(\alpha x\right)J^{\prime}_{n}\left(\alpha x\right)e^{-\rho^{2}x^{2}}.\label{eq:RKAP5:Bessel_int_ident-1}
\end{multline}

Now we employ recursion relations of the Bessel functions in order
to write 
\[
J^{\prime2}_{n}\left(z\right)=\frac{n^{2}}{z^{2}}J^{2}_{n}\left(z\right)-J_{n-1}\left(z\right)J_{n+1}\left(z\right).
\]
Combining this result with (\ref{eq:RKAP5:Bessel_int_ident-1}) we
obtain a second identity, 
\begin{equation}
\int^{\infty}_{0}dx\,x^{2}J_{n}\left(\alpha x\right)J^{\prime}_{n}\left(\alpha x\right)e^{-\rho^{2}x^{2}}=\frac{\rho^{2}}{\alpha}\int^{\infty}_{0}dx\,x^{3}J_{n-1}\left(\alpha x\right)J_{n+1}\left(\alpha x\right)e^{-\rho^{2}x^{2}}.\label{eq:RKAP5:Bessel_int_ident-2}
\end{equation}

Now, \citet{Gaelzer+16/06} defined the function 
\begin{multline*}
\mathcal{Y}^{\left(\alpha,\beta\right)}_{n,\kappa}\left(\varrho,\varsigma\right)=\frac{2}{\pi^{1/2}\kappa^{1/2+\beta}\varrho}\frac{\Gamma\left(\lambda-1\right)}{\Gamma\left(\sigma-\nicefrac{3}{2}\right)}\\
\times\int^{\infty}_{0}dx\,\int^{\infty}_{-\infty}ds\,\frac{x^{3}J_{n-1}\left(\nu x\right)J_{n+1}\left(\nu x\right)}{s-\varsigma}\left(1+\frac{x^{2}}{\kappa}+\frac{s^{2}}{\kappa}\right)^{-\left(\lambda-1\right)},
\end{multline*}
where $\sigma=\kappa+\alpha$, $\lambda=\sigma+\beta$ and $\nu=\sqrt{2\varrho}$\@.
If we first employ identity (\ref{eq:RKAP5:Superstatistics-ident})
in order to write 
\[
\left(1+\frac{x^{2}}{\kappa}+\frac{s^{2}}{\kappa}\right)^{-\left(\lambda-1\right)}=\frac{1}{\Gamma\left(\lambda-1\right)}\int^{\infty}_{0}dt\,t^{\lambda-2}e^{-t}e^{-tx^{2}/\kappa}e^{-ts^{2}/\kappa},
\]
we can employ identity (\ref{eq:RKAP5:Bessel_int_ident-2}) to write
\[
\int^{\infty}_{0}dx\,x^{3}J_{n-1}\left(\nu x\right)J_{n+1}\left(\nu x\right)e^{-tx^{2}/\kappa}=\frac{\kappa\varrho}{t}\frac{\partial\hphantom{\varrho}}{\partial\varrho}\int^{\infty}_{0}dx\,xJ^{2}_{n}\left(\nu x\right)e^{-tx^{2}/\kappa}.
\]
Then, by rearranging the order of integrals, evaluating the $t$-integral
and referring to (\ref{eq:SKDL-2}), we obtain the desired result.

\bibliographystyle{jpp}

\bibliography{Oblique_bRKd_LyX}

\end{document}